\documentclass[fleqn,usenatbib]{mnras}
\PassOptionsToPackage{table}{xcolor}
\usepackage{graphicx} 
\usepackage{amsmath}      
\usepackage{amssymb}      
\usepackage{multicol}        
\usepackage{pdflscape}	
\usepackage{orcidlink}  
\usepackage{xcolor}
\usepackage[normalem]{ulem}
\usepackage{tikz}
\usetikzlibrary{decorations.markings}
\usepackage{xspace}
\usepackage{diagbox}

\usepackage[T1]{fontenc}
\usepackage{ae,aecompl}
\usepackage{newtxtext,newtxmath} 

\newcommand{\orcid}[1]{\href{https://orcid.orffi#1}{\textcolor[HTML]{A6CE39}{\aiOrcid}}}

\newcommand{\Stwentythree}{\href{https://doi.org/10.1093/mnras/stac3263}{S23}\xspace}

\begin{document}

\author[S. DeLaurentiis et al.]{
Stanislav DeLaurentiis$^{1}$\orcidlink{0000-0002-8922-825X}\thanks{Contact e-mail: \href{mailto:sod2112@columbia.edu}{sod2112@columbia.edu}}, Zolt\'an Haiman$^{2,1,3}$\orcidlink{0000-0003-3633-5403}\thanks{Contact e-mail: \href{mailto:zh2007@columbia.edu}{zh2007@columbia.edu}}, Magdalena Siwek$^{4}$\orcidlink{0000-0002-1530-9778}\thanks{Contact e-mail: \href{mailto:mss2334@columbia.edu}{mss2334@columbia.edu}}, 
\\
$^{1}$Department of Astronomy, Columbia University, 550 W. 120th Street, New York, NY 10027, USA\\
$^{2}$Institute of Science and Technology Austria (ISTA), Am Campus 1, 3400 Klosterneuburg, Austria\\
$^{3}$Department of Physics, Columbia University, 550 W. 120th Street, New York, NY 10027, USA\\
$^{4}$Center for Cosmology and Particle Physics, Department of Physics, New York University, 726 Broadway, New York, NY 10003, USA\\
}

\title{Preferential accretion onto eccentric and unequal binary black holes}

\date{}

\pubyear{2024}

\label{firstpage}
\pagerange{\pageref{firstpage}--\pageref{lastpage}}
\maketitle

\begin{abstract}
    Supermassive binary black holes (SMBBHs) are expected to be surrounded by circumbinary disks (CBDs) which affect the binary through gravitational forces and accretion. It has been reported that the binary can experience ``preferential accretion'' where one black hole (BH) out-accretes the other for hundreds of orbits, but this asymmetry has yet to be fully described or understood.  In this work, we utilize a suite of 80 SMBBH hydrodynamical simulations with varying mass ratios ($q_b$) and eccentricity ($e_b$) in order to robustly delineate the behavior of preferential accretion, determine its relationship to the structure of the CBD, and study its observational consequences. We characterize the accretion-rate ratio $\lambda(t) \equiv \dot{M}_2(t)/\dot{M}_1(t)$ and the mass-ratio rate of change $\dot{q}_b \equiv d/dt(M_2/M_1)$ across the suite. We confirm that the secondary tends to out-accrete the primary  ($\lambda \geq 1$), and find this preference to be strongest for low-$e_b$, low-$q_b$ binaries and increasingly time-variable toward high $e_b$. We also find that (i) the time-variability of $\lambda$ tracks the precession of the CBD, (ii) there can be sub- and super-Eddington accretion in a single binary, and (iii) the gas-driven approach toward equal mass becomes particularly slow for highly eccentric, high $q_b$ binaries, suggesting that some binaries may not reach $q_b=1$ within the $30\,\mathrm{Myr}$ lifetime of a quasar and therefore allowing LISA to constrain the accretion history of SMBBHs. Our findings also suggest that periodically flickering jets are a potential observable signature of many binaries.
    
\end{abstract}

\section{Introduction}

Cosmic structure forms hierarchically and galaxy mergers are expected to result in gravitationally bound supermassive binary black holes (SMBBHs; \citealt{white_rees_78, begelman_blandford_rees_80}). During a galactic merger, it is also expected that the inter-stellar medium from the proto-galaxies are funneled to the galactic center \citep{barnes_hernquist_92}. This creates a reservoir of material which, due to the conservation of angular momentum and the binary's gravitational potential, forms a circumbinary disk (CBD). Many aspects of the CBD have been well studied in the literature. It has been shown that the binary can influence the CBD, creating eccentric structure and precessing eigenmodes, \citep{lubow_91, whitehurst_94, nelson_2003, Goodchild_2006, MacFadyen_08, paardekooper_2008, kley_2008, shi_2012, miranda_munoz_lai_2017, Thun_2017, munoz_lithwick_2020, Lubow_2022, siwek_prefacc} and that the CBD can in turn influence the binary, changing both its semi-major axis and eccentricity \citep{roedig_2011, farris_2014, Moody_19, munoz_miranda_lai_19, Tiede_2020, zrake_2021, dorazio_2013, siwek_cbdorbevol, dorazio_duffel}, as well as exciting precession of the binary itself \citep{tiede_24_diskinducedprecession, dittmann_prec, calcino_prec}. However, crucially, the binary also accretes from the circumbinary disk, influencing both the mass-ratio and the associated light-curves and spectra of the system (e.g.\ \citealt{dorazio_2013, dorazio_charisi, siwek_prefacc, ryan_2022, farris_2014, tiede_dorazio_2025_hotcoldmulticomponentaccretion}).

It has been a widely known result of CBD studies that the accretion on to the binary ($\dot{M}_{\rm{b}} = \dot{M}_1 + \dot{M}_2$) is variable. Near equal-mass ratio ($q_b \equiv M_2/M_1 \gtrsim 0.7$) circular binaries display a sawtooth pattern with a period of about $5$ binary orbital periods ($\tau_{\rm{b}}$) potentially due to the presence of an $\rm{m}=1$ over-density traveling at the inner edge of the CBD \citep{MacFadyen_08, miranda_munoz_lai_2017, munoz_miranda_lai_19, ryan_2022}. Eccentric and unequal-mass binaries also display variability on the order of a binary period \citep{dorazio_2013, farris_2014, miranda_munoz_lai_2017, munoz_miranda_lai_19, duffell_dorazio_2020, ryan_2022}. In order to study how the binary accretes material from the circumbinary disk, \citet{Tiede_22} tracked the paths of passive tracer particles in a two-dimensional grid-based hydrodynamical code (DISCO). They determined that much of the gas is viscously transported from the outer disk before adopting a nearly ballistic trajectory and ultimately accreting on to one of the BHs. They also delineate an ``accretion horizon'' at $r \approx 1.05 \, a_b$ (the binary semi-major axis), a radius past which any material entering is accreted.

Studies have also pointed out the disparity between the accretion rates of the primary and secondary components of the SMBBH (e.g.\ \citealt{Dorazio+2013,farris_2014,miranda_munoz_lai_2017, siwek_prefacc, Dorazio_24_pref_mention}). Namely, the secondary tends to accrete at a greater rate in what has become known as ``preferential accretion'', pushing the system toward equal mass ratio ($q_b = 1$; \citealt{dorazio_2013, farris_2014, duffell_dorazio_2020, miranda_munoz_lai_2017, siwek_prefacc}). While systems with $q_b = 1$ generally accrete at equal time-averaged rates, suggesting a stable equilibrium at $q_b = 1$, studies have reported that at certain eccentricities (e.g.\ $e_b = 0.5$ and $e_b = 0.6$) the binary can undergo transient ``symmetry breaking'' (see Figure 7 of \citealt{munoz_miranda_lai_19}). In these instances, one black hole temporarily accretes more than its companion before its accretion rate is suppressed, allowing the other to experience an enhanced accretion-rate episode.

This process was explored in \citealt{siwek_prefacc}, hereafter \Stwentythree, using a suite of $80$ two-dimensional (2D) hydrodynamical simulations of SMBBHs with CBDs, sampling different values of the parameters $q_b$ and $e_b$. \Stwentythree reported time-averaged values of $\lambda = \dot{M}_2/\dot{M}_1$, the ratio of the secondary's accretion rate to the primary's, for their simulation suite and displayed snapshots of the corresponding CBD. They found that low-eccentricity binaries display larger values of $\lambda$ than high-eccentricity binaries ($e_b > 0.6$). Further, the \textit{flip} in preferential accretion, the switch in which BH (primary or secondary) momentarily accretes at the higher rate, was suggested to be associated with a unique disk behavior they called ``forced precession''. \citet{DeLaurentiis24} also discussed the preferential accretion rate of the binary, suggesting that it can be understood in geometric terms, through the distance from the individual BHs to the nearest point of the CBD.

The protoplanetary-disk community has run two- and three-dimensional smoothed-particle hydrodynamic (SPH) simulations of CBDs to understand the accretion onto the primary and secondary \citep{Gunther_Kley_02, Ochi_05, Young_15}. To date, in simulations run for up to $\sim 100$ binary orbital periods ($\tau_b$), they have found that the primary can out-accrete the secondary due to streamlines that, despite entering the Lagrange point L2 near the secondary, carry enough angular momentum to flow back to the primary \citep{Ochi_05, Young_15}. Similar findings are reported by \citet{Tiede_22}.

Despite these efforts we are still yet to fully characterize and understand the rich detail of accretion onto the individual components of a binary in a CBD system. Thus, we build on \Stwentythree and use the larger simulation suite of \citealt{siwek_cbdorbevol} to understand fundamental gas dynamics near BH binaries across different $(q_b, e_b)$, with implications for both their long-term orbital evolution and observational signatures, including light-curves and spectra.

In this study we report the largest binary parameter sweep study on ``preferential accretion'' to date. Further, we characterize the accretion behavior by drawing on linear theory of eccentric circumbinary disks and applying a few targeted numerical techniques: Fourier-based period extraction of the accretion rate ratio, comparison against the individual black hole Eddington threshold, and a coupled gas-plus-gravitational-wave integration of the evolution of the binary mass ratio and eccentricity to study its long-term behavior.

In \autoref{sec:methods} we briefly discuss the technical details of the simulations, the key concepts from linear theory we draw upon, and the numerical techniques we employ. In \autoref{sec:results} we present our findings, detailing the temporal behavior of preferential accretion and mass-ratio evolution and describing how each varies with $q_b$ and $e_b$. In \autoref{sec:observations} we discuss the consequences of these findings for observations: in particular, jets that periodically switch on and off (``flickering'' jets) and a population of SMBBHs with $q_b \neq 1$. In \autoref{sec:conclusion} we summarize our key findings and discuss next steps.

\section{Analytic tools and numerical methods}\label{sec:methods}
In the following section we briefly describe the setup of the \citet{siwek_prefacc, siwek_cbdorbevol} simulations, the concepts from linear disk theory we employ, and the numerical tools we leverage.
\subsection{Simulation setup}
We briefly describe the setup of the simulations of interest and refer readers to \Stwentythree for a more thorough discussion.

\Stwentythree performed 2D hydrodynamical simulations of binary black holes embedded in a finite, locally isothermal disk with Mach number $\mathcal{M}=10$, an $\alpha$-viscosity of $\alpha=0.1$, and an aspect ratio of $h\equiv H/r=0.1$. The disk is initialized with a power-law surface-density profile and a corresponding temperature profile; the inner edge of the disk sits at $2a_b$ and the outer edge at $\approx 50a_b$. The binary-disk system spans a computational grid of $300 a_b \times 300a_b$ with open boundary conditions, allowing the disk to settle into a quasi-steady state.

The binary is modeled as two sink particles with a mass-ratio $q_b \equiv M_2/M_1 \leq 1$ and radii $r_{\rm{s}} = 0.03\, a_{\rm{b}}$ (the same sink radius is adopted for both BHs regardless of $q_b$; this is a numerical choice rather than a physical scale), moving on a fixed Keplerian orbit of eccentricity $e_{\rm{b}}$. For each gas cell lying inside a sink, i.e.\ where $r_{ij} \leq r_{\rm{s}}$ (with $r_{ij}$ the radial distance from the j$^{\rm{th}}$ sink particle to the i$^{\rm{th}}$ gas cell), the fraction of its gas accreted by that sink particle at each time step is $\gamma_0 \left( 1 - \frac{r_{ij}}{r_{\rm{s}}}\right)^2$. In addition to mass, the sink also accretes the gas' linear momentum from the gas cells.

The simulations, conducted with the moving-mesh code AREPO \citep{Springel_arepo_10}, use Voronoi tessellations to generate a grid of cells and explore a wide parameter space spanning $q_b \in \{0.1, 0.2, 0.3, 0.4, 0.5, 0.6, 0.7, 0.8, 0.9, 1.0\}$ and $e_b \in \{0.0, 0.1, 0.2, 0.3, 0.4, 0.5, 0.6, 0.8\}$. The simulations were run for 10,000 binary orbits, with surface-density snapshots recorded at apocenter every $10$ binary orbits. Accretion rates are recorded independently at a much higher cadence, with a time-step of $\approx 0.01$ orbits.

\begin{figure}
    \centering
    \includegraphics[width=\linewidth]{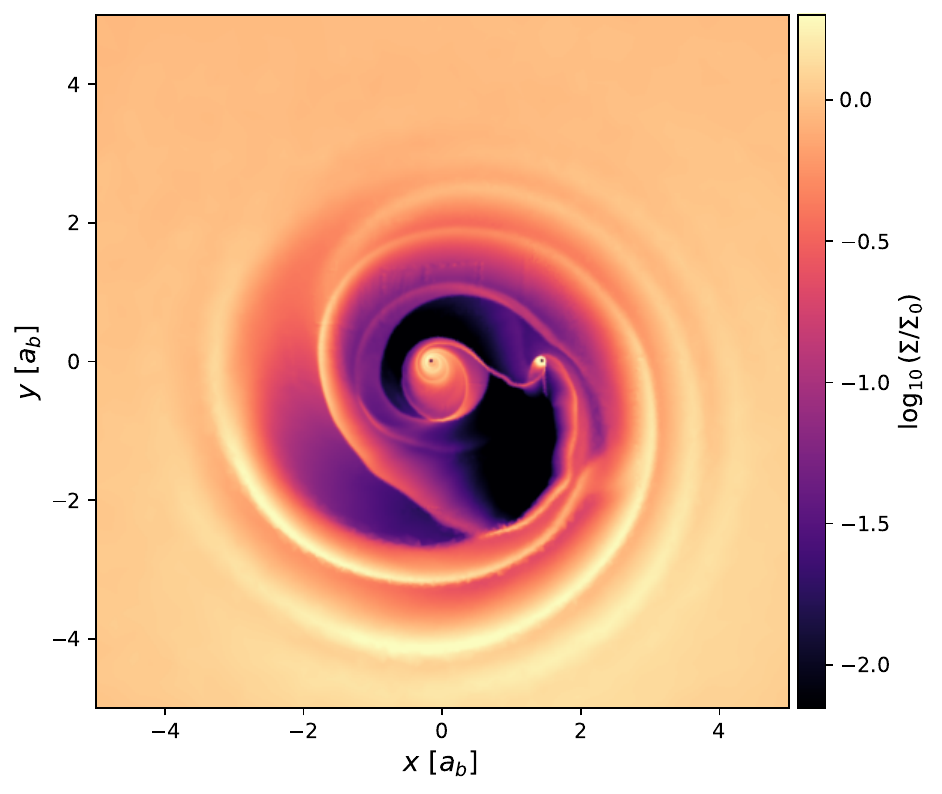}
    \caption{Representative instantaneous surface-density map from the simulation suite, for the $(e_b, q_b) = (0.6, 0.1)$ binary at apocenter (after ${\approx}\,4000$ binary orbits, well past the initial transient). The color scale is the surface density on a logarithmic scale, normalized to a fiducial outer-disk value $\Sigma_0$. The binary clears an eccentric, lopsided cavity; the two black holes (the bright, compact minidiscs near the center) accrete through narrow streams that cross the cavity and sit at unequal distances from the cavity wall. Axes are in units of the binary semi-major axis $a_b$, centered on the binary's center of mass.}
    \label{fig:cavity_snapshot}
\end{figure}

\autoref{fig:cavity_snapshot} shows a representative surface-density snapshot from the suite, for the $(e_b, q_b) = (0.6, 0.1)$ binary at apocenter. It illustrates the features that recur across the eccentric simulations and that we draw on throughout this work: a low-density inner \textit{cavity} cleared by the binary; the pronounced lopsidedness of that cavity, whose wall lies much closer to the binary on one side than on the other; the compact minidiscs (the bright points near the center) that form around each black hole; and the narrow gas streams that penetrate the cavity and feed those minidiscs. Because the cavity is lopsided, the two black holes generally sit at different distances from the cavity wall, a geometric asymmetry that we will argue is central to preferential accretion and its time variability (\S\ref{sec:preliminary_analysis}).

\subsection{Disk theory}

In order to illuminate the relationship between the accretion of the binary and the precession and eccentricity of the CBD we must first characterize the time-variable attributes of the disk. Namely, we focus on the eccentricity and precession of the disk. By the disk eccentricity (or ``gas eccentricity'') we mean the mass-weighted orbital eccentricity of the gas, obtained from the local eccentricity vector
\begin{equation}\label{eqn:ecc_vector}
    \mathbf{e}(\mathbf{r}) = \frac{\mathbf{v}\times(\mathbf{r}\times\mathbf{v})}{GM_b} - \hat{\mathbf{r}}
\end{equation}
of each fluid element with velocity $v$ at distance $r$ from the binary's center of mass with mass $M_b$, in practice we characterize it through the shape of the cavity (its inner edge), as described below.

Disk eccentricity within CBDs is expected to grow through mechanisms such as eccentric Lindblad resonances (ELRs) or spiral shock pumping at the cavity edge \citep{lubow_91, whitehurst_94, paardekooper_2008, kley_2008, shi_2012} or orbital instabilities \citep{dorazio_2013, mcwilliams_instability}. While ELRs, spiral shocks, and orbital instabilities promote disk eccentricity growth \citep{lubow_91, shi_2012}, viscous damping acts to suppress it \citep{Goodchild_2006}. Previous 2D simulations have found steady-state eccentricity profiles, indicating that these competing effects can reach equilibrium \citep{miranda_munoz_lai_2017, siwek_prefacc}.

Further, 2D and 3D hydrodynamical simulations have found significant disk eccentricity near the inner edge (the cavity), with the eccentricity declining outward \citep{MacFadyen_08, miranda_munoz_lai_2017, siwek_prefacc, ragusa_lynch}. Similar trends have been seen in magneto-hydrodynamical simulations \citep{shi_2012}, suggesting that disk eccentricity is a robust characteristic of the inner regions of eccentric CBDs.

Additionally, simulations have demonstrated that CBDs can precess \citep{nelson_2003, shi_2012, miranda_munoz_lai_2017, Thun_2017, siwek_prefacc}, with precession frequencies attributed to the eigenmodes of a Schrödinger-like equation for eccentricity evolution \citep{Goodchild_2006, shi_2012, Teyssandier_2016, Lee_2019, munoz_lithwick_2020, Lubow_2022}.

A robust study of the eccentricity and precession of both the bulk of the disk and the cavity in hydrodynamical simulations was conducted by DeLaurentiis \& Rafikov (in preparation). Utilizing the same suite of simulations as this paper, they delineate the shape of the non-linear inner edge of the circumbinary disk, the cavity, by extracting the dominant Fourier modes of the associated isodensity contour.

While alternative explanations have been proposed \citep{Artimowicz_83_og_polish}, it has widely been assumed that the preferential accretion of the binary is related to the relative closeness of the components to the CBD's inner edge \citep{Dorazio+2013,Rafikov_16_accretion}. In order to test this dependence, we utilize the results for the cavity shape and its precession period from DeLaurentiis \& Rafikov (in preparation) to build a robust time-dependent geometric picture of the binary in the cavity.

\subsection{Numerical techniques}
As discussed earlier, in order to highlight the link between accretion and the precession of the CBD, we are interested in understanding both as time-dependent quantities. The orientation of the CBD is inherently time-dependent due to its apsidal precession\footnote{While the eccentricity and semi-major axis of the cavity are strictly time-dependent, for most systems we find that the shape of the cavity achieves a steady state and its time-dependence is dominated by its apsidal precession alone.}. Since our simulation runs output snapshots every $10$ orbits, the time-series associated with our CBD is constrained to a $10\,\tau_b$ cadence, where $\tau_b$ denotes the binary orbital period.

The accretion rate is likewise time-dependent. For each sink, the simulation tracks the mass accreted $\Delta m$ at time-step $t$, with a maximum cadence of $0.01\,\tau_b$. As we are interested in comparing the accretion rate to the precession of the CBD, we boxcar-average the high-cadence accretion-rate data to yield a time-series that matches the $10\,\tau_b$ cadence of the CBD time-series. We do this by summing the instantaneous mass accreted over non-overlapping $10\,\tau_b$ windows and dividing by the window size. We emphasize that this operation is a \textit{smoothing} of $\dot{M}$ over $10\,\tau_b$ scales rather than a strict down-sampling: the CBD snapshots are recorded instantaneously at apocenter every $10\,\tau_b$, whereas $\dot{M}$ is averaged over the intervening window. Because the modulation we compare (the precession-paced variability of $\lambda$ and the cavity orientation) varies on the much longer precession timescale ($\sim 10^2$--$10^3\,\tau_b$), this difference in sub-window treatment does not affect the periods or correlations we report. After we transform the accretion rates for each BH into the lower cadence of $10$ orbits, we construct our preferential accretion quantity $\lambda(t) \equiv \frac{\dot{M}_2(t)}{\dot{M}_1(t)}$, the ratio of the accretion rate of the secondary to the primary.

We also report the rate of change of the mass-ratio $\dot{q}_b(t)$. We define it as
\begin{equation}\label{eqn:q_dot}
    \dot{q}_b = \frac{[1+q_b(t)][\lambda(q_b) - q_b(t)]}{1 + \lambda(q_b)}\frac{\dot{M}_1 +\dot{M}_2}{M_1 + M_2}
\end{equation}
where
\begin{equation}
    q_b(t) \equiv \frac{M_2 (t)}{M_1 (t)}
\end{equation}
and the mass of each BH
\begin{equation}
    M_j(t) = M_j(t=0) + \sum_{t'=0}^{t} \Delta m(t')
\end{equation}
is simply the running sum. Again, we note that all time-variable inputs are first transformed to be of $10$-orbit cadence via the aforementioned procedure.

\section{Results}\label{sec:results}
In the following sections we report our key results. Namely, in \autoref{sec:pref_acc} we detail the behavior of $\lambda(t)$ as a function of $e_b$ and $q_b$ and report its mean value and, if time variable, its period. In \autoref{sec:preliminary_analysis}, we speculate about the CBD's effect on the accretion behavior. In \autoref{sec:mass_ratio} we calculate the corresponding rate of change of the mass ratio, and highlight an instance of the binary accreting away from equal mass ($q_b = 1$).

\subsection{Preferential accretion}\label{sec:pref_acc}

\begin{figure*}
    \centering
    \includegraphics[width=\textwidth,height=0.85\textheight,keepaspectratio]{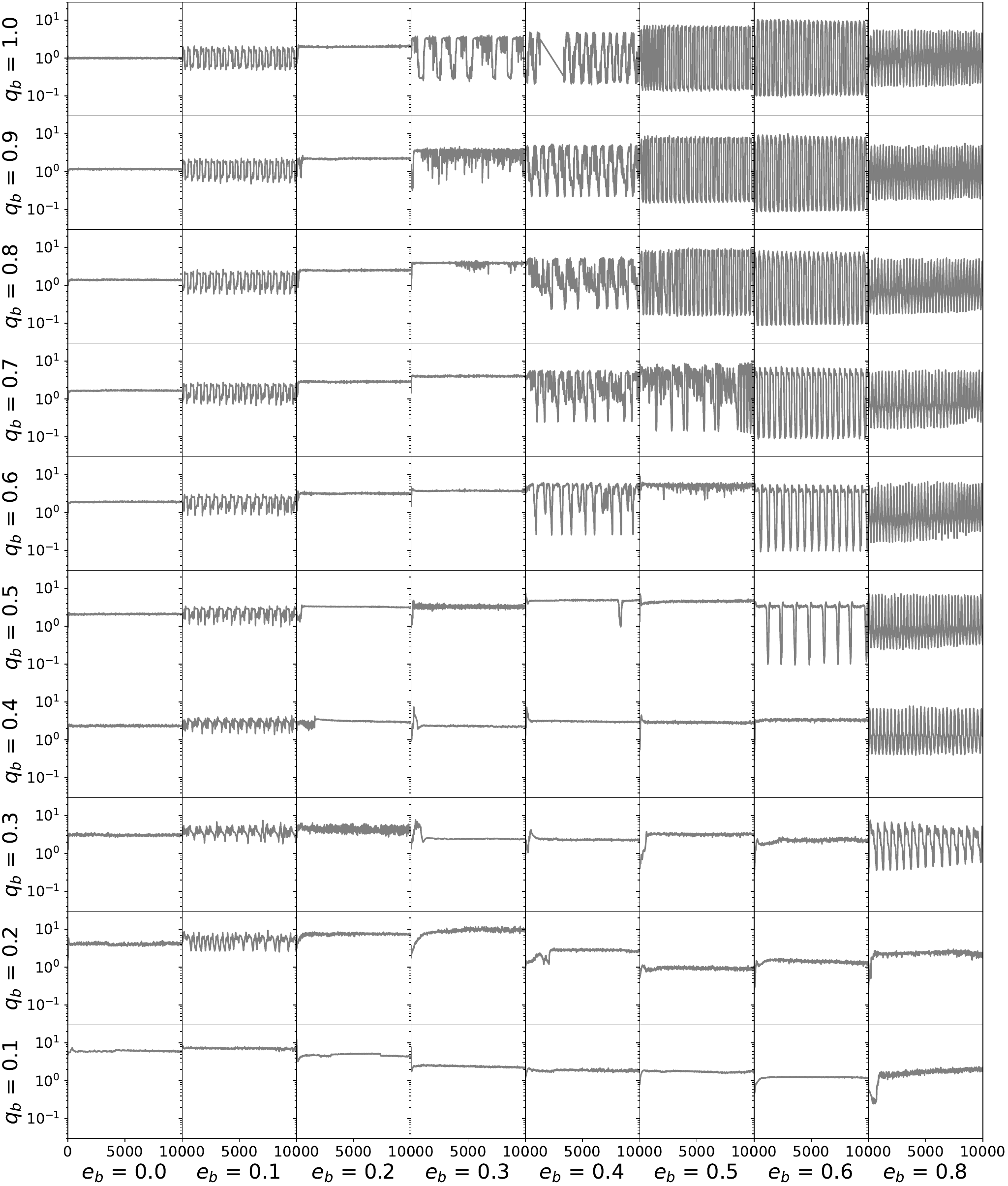}
    \caption{The accretion-rate ratio $\lambda(t) \equiv \dot{M}_2(t) / \dot{M}_1(t)$ for the entire 80-simulation suite. Each panel is a unique $(e_b, q_b)$ simulation, with $\lambda(t)$ on the y-axis and time on the x-axis (in units of binary orbital period $\tau_b$). The panel's position on the larger grid signifies its binary parameters: $e_b$ is constant along columns and increases left-to-right; $q_b$ is constant along rows and increases bottom-to-top. $\lambda(t)$ divides cleanly into time-stable and time-varying regimes across the suite, as illustrated in Table~\ref{tab:stable_varying_grid}.}
    \label{fig:lambda_full}
\end{figure*}

In this section, we first give a detailed account of the characteristics of preferential accretion as measured in our simulation suite (\ref{subsec:prefer-describe}), followed by a discussion on possible physical interpretations (\ref{sec:preliminary_analysis}).

\subsubsection{Characteristics of preferential accretion} \label{subsec:prefer-describe}

In \autoref{fig:lambda_full} we show our $\lambda(t)$ time-series for the entirety of our simulation suite. The figure is structured such that each panel represents a unique simulation. The rows show mass-ratio $q_b$, whereas the columns show eccentricity $e_b$. The left-most columns are the least eccentric binaries, and the upper-most rows are the most equal-mass binaries.

Before turning to the ratio $\lambda$, we note that the \textit{total} accretion rate $\dot{M}_b = \dot{M}_1 + \dot{M}_2$ is comparable across the suite: averaged over the post-transient window it varies by only ${\approx}\,10\%$ ($1\sigma$) about its median, with a factor of ${\approx}\,2$ between the most extreme cells. Because the binaries grow at a similar total rate, and because we later rescale $\dot{M}_b$ to Eddington units (\S\ref{sec:observations}), the variation in $\lambda$ that we focus on below reflects how a roughly fixed total supply is partitioned between the two BHs, rather than differences in the total gas supply.

A striking feature of \autoref{fig:lambda_full} is the broad division of $\lambda(t)$ into either approximately constant or time-varying behavior. Many low-$e_b$, low-$q_b$ binaries are time-stable. Though some experience a sharp change in behavior in the first $2000$ orbits associated with the expected initial numerical disk-instability transient (eg. \citealt{moriwaki_04}), they soon settle to a near-constant value. This time-stable behavior is exemplified by the $(e_b, q_b) = (0.3, 0.3)$ binary\footnote{Some binaries (e.g.\ $(e_b, q_b) = (0.0, 0.3)$) experience noticeable numerical noise around the constant value, but the dichotomy between time-varying and time-stable $\lambda(t)$ remains clear.}. Others show time-variable behavior, with oscillations in $\lambda(t)$ spanning as much as two orders of magnitude, such as $(e_b, q_b) = (0.6, 1.0)$.

\renewcommand{\arraystretch}{1.5} 
\begin{table}
\centering
\begin{tabular}{|c|c|c|c|c|c|c|c|c|}
    \hline
    \textbf{1.0} &
    \cellcolor{red}\textcolor{white}{S} &
    \cellcolor{blue}\textcolor{white}{V} &
    \cellcolor{red}\textcolor{white}{S} &
    \cellcolor{blue}\textcolor{white}{V} &
    \cellcolor{blue}\textcolor{white}{V} &
    \cellcolor{blue}\textcolor{white}{V} &
    \cellcolor{blue}\textcolor{white}{V} &
    \cellcolor{blue}\textcolor{white}{V} \\
    \hline
    \textbf{0.9} &
    \cellcolor{red}\textcolor{white}{S} &
    \cellcolor{blue}\textcolor{white}{V} &
    \cellcolor{red}\textcolor{white}{S} &
    \cellcolor{red}\textcolor{white}{S} &
    \cellcolor{blue}\textcolor{white}{V} &
    \cellcolor{blue}\textcolor{white}{V} &
    \cellcolor{blue}\textcolor{white}{V} &
    \cellcolor{blue}\textcolor{white}{V} \\
    \hline
    \textbf{0.8} &
    \cellcolor{red}\textcolor{white}{S} &
    \cellcolor{blue}\textcolor{white}{V} &
    \cellcolor{red}\textcolor{white}{S} &
    \cellcolor{red}\textcolor{white}{S} &
    \cellcolor{blue}\textcolor{white}{V} &
    \cellcolor{blue}\textcolor{white}{V} &
    \cellcolor{blue}\textcolor{white}{V} &
    \cellcolor{blue}\textcolor{white}{V} \\
    \hline
    \textbf{0.7} &
    \cellcolor{red}\textcolor{white}{S} &
    \cellcolor{blue}\textcolor{white}{V} &
    \cellcolor{red}\textcolor{white}{S} &
    \cellcolor{red}\textcolor{white}{S} &
    \cellcolor{blue}\textcolor{white}{V} &
    \cellcolor{blue}\textcolor{white}{V} &
    \cellcolor{blue}\textcolor{white}{V} &
    \cellcolor{blue}\textcolor{white}{V} \\
    \hline
    \textbf{0.6} &
    \cellcolor{red}\textcolor{white}{S} &
    \cellcolor{blue}\textcolor{white}{V} &
    \cellcolor{red}\textcolor{white}{S} &
    \cellcolor{red}\textcolor{white}{S} &
    \cellcolor{blue}\textcolor{white}{V} &
    \cellcolor{red}\textcolor{white}{S} &
    \cellcolor{blue}\textcolor{white}{V} &
    \cellcolor{blue}\textcolor{white}{V} \\
    \hline
    \textbf{0.5} &
    \cellcolor{red}\textcolor{white}{S} &
    \cellcolor{blue}\textcolor{white}{V} &
    \cellcolor{red}\textcolor{white}{S} &
    \cellcolor{red}\textcolor{white}{S} &
    \cellcolor{red}\textcolor{white}{S} &
    \cellcolor{red}\textcolor{white}{S} &
    \cellcolor{blue}\textcolor{white}{V} & \cellcolor{blue}\textcolor{white}{V} \\
    \hline
    \textbf{0.4} &
    \cellcolor{red}\textcolor{white}{S} &
    \cellcolor{blue}\textcolor{white}{V} &
    \cellcolor{red}\textcolor{white}{S} &
    \cellcolor{red}\textcolor{white}{S} &
    \cellcolor{red}\textcolor{white}{S} &
    \cellcolor{red}\textcolor{white}{S} &
    \cellcolor{red}\textcolor{white}{S} &
    \cellcolor{blue}\textcolor{white}{V} \\
    \hline
    \textbf{0.3} &
    \cellcolor{red}\textcolor{white}{S} &
    \cellcolor{blue}\textcolor{white}{V} &
    \cellcolor{red}\textcolor{white}{S} &
    \cellcolor{red}\textcolor{white}{S} &
    \cellcolor{red}\textcolor{white}{S} &
    \cellcolor{red}\textcolor{white}{S} &
    \cellcolor{red}\textcolor{white}{S} &
    \cellcolor{blue}\textcolor{white}{V} \\
    \hline
    \textbf{0.2} &
    \cellcolor{red}\textcolor{white}{S} &
    \cellcolor{blue}\textcolor{white}{V} &
    \cellcolor{red}\textcolor{white}{S} &
    \cellcolor{red}\textcolor{white}{S} &
    \cellcolor{red}\textcolor{white}{S} &
    \cellcolor{red}\textcolor{white}{S} &
    \cellcolor{red}\textcolor{white}{S} &
    \cellcolor{red}\textcolor{white}{S} \\
    \hline
    \textbf{0.1} &
    \cellcolor{red}\textcolor{white}{S} &
    \cellcolor{red}\textcolor{white}{S} &
    \cellcolor{red}\textcolor{white}{S} &
    \cellcolor{red}\textcolor{white}{S} &
    \cellcolor{red}\textcolor{white}{S} &
    \cellcolor{red}\textcolor{white}{S} &
    \cellcolor{red}\textcolor{white}{S} &
    \cellcolor{red}\textcolor{white}{S} \\
    \hline
    \multicolumn{1}{|c|}{\diagbox[dir=NE,height=2.2\line]{$q_b$}{$e_b$}} & \textbf{0.0} & \textbf{0.1} & \textbf{0.2} & \textbf{0.3} & \textbf{0.4} & \textbf{0.5} & \textbf{0.6} & \textbf{0.8} \\
    \hline
\end{tabular}
\caption{Grid showing whether $\lambda(t)$ for a given binary is time-stable (S, red) or time-varying (V, blue). This time-stable/time-varying split closely matches the locked/precessing partition of the CBD in \autoref{tab:locked_precessing_grid}.}
\label{tab:stable_varying_grid}
\end{table}

\begin{table}
\centering
\begin{tabular}{|c|c|c|c|c|c|c|c|c|}
    \hline
    \textbf{1.0} &
    \cellcolor{blue}\textcolor{white}{P} &
    \cellcolor{blue}\textcolor{white}{P} &
    \cellcolor{red}\textcolor{white}{L} &
    \cellcolor{blue}\textcolor{white}{P} &
    \cellcolor{blue}\textcolor{white}{P} &
    \cellcolor{blue}\textcolor{white}{P} &
    \cellcolor{blue}\textcolor{white}{P} &
    \cellcolor{blue}\textcolor{white}{P} \\
    \hline
    \textbf{0.9} &
    \cellcolor{blue}\textcolor{white}{P} &
    \cellcolor{blue}\textcolor{white}{P} &
    \cellcolor{red}\textcolor{white}{L} &
    \cellcolor{red}\textcolor{white}{L} &
    \cellcolor{blue}\textcolor{white}{P} &
    \cellcolor{blue}\textcolor{white}{P} &
    \cellcolor{blue}\textcolor{white}{P} &
    \cellcolor{blue}\textcolor{white}{P} \\
    \hline
    \textbf{0.8} &
    \cellcolor{blue}\textcolor{white}{P} &
    \cellcolor{blue}\textcolor{white}{P} &
    \cellcolor{red}\textcolor{white}{L} &
    \cellcolor{red}\textcolor{white}{L} &
    \cellcolor{blue}\textcolor{white}{P} &
    \cellcolor{blue}\textcolor{white}{P} &
    \cellcolor{blue}\textcolor{white}{P} &
    \cellcolor{blue}\textcolor{white}{P} \\
    \hline
    \textbf{0.7} &
    \cellcolor{blue}\textcolor{white}{P} &
    \cellcolor{blue}\textcolor{white}{P} &
    \cellcolor{red}\textcolor{white}{L} &
    \cellcolor{red}\textcolor{white}{L} &
    \cellcolor{blue}\textcolor{white}{P} &
    \cellcolor{blue}\textcolor{white}{P} &
    \cellcolor{blue}\textcolor{white}{P} &
    \cellcolor{blue}\textcolor{white}{P} \\
    \hline
    \textbf{0.6} &
    \cellcolor{blue}\textcolor{white}{P} &
    \cellcolor{blue}\textcolor{white}{P} &
    \cellcolor{red}\textcolor{white}{L} &
    \cellcolor{red}\textcolor{white}{L} &
    \cellcolor{blue}\textcolor{white}{P} &
    \cellcolor{red}\textcolor{white}{L} &
    \cellcolor{blue}\textcolor{white}{P} &
    \cellcolor{blue}\textcolor{white}{P} \\
    \hline
    \textbf{0.5} &
    \cellcolor{blue}\textcolor{white}{P} &
    \cellcolor{blue}\textcolor{white}{P} &
    \cellcolor{red}\textcolor{white}{L} &
    \cellcolor{red}\textcolor{white}{L} &
    \cellcolor{red}\textcolor{white}{L} &
    \cellcolor{red}\textcolor{white}{L} &
    \cellcolor{blue}\textcolor{white}{P} & \cellcolor{blue}\textcolor{white}{P} \\
    \hline
    \textbf{0.4} &
    \cellcolor{blue}\textcolor{white}{P} &
    \cellcolor{blue}\textcolor{white}{P} &
    \cellcolor{red}\textcolor{white}{L} &
    \cellcolor{red}\textcolor{white}{L} &
    \cellcolor{red}\textcolor{white}{L} &
    \cellcolor{red}\textcolor{white}{L} &
    \cellcolor{red}\textcolor{white}{L} &
    \cellcolor{blue}\textcolor{white}{P} \\
    \hline
    \textbf{0.3} &
    \cellcolor{blue}\textcolor{white}{P} &
    \cellcolor{blue}\textcolor{white}{P} &
    \cellcolor{red}\textcolor{white}{L} &
    \cellcolor{red}\textcolor{white}{L} &
    \cellcolor{red}\textcolor{white}{L} &
    \cellcolor{red}\textcolor{white}{L} &
    \cellcolor{red}\textcolor{white}{L} &
    \cellcolor{blue}\textcolor{white}{P} \\
    \hline
    \textbf{0.2} &
    \cellcolor{blue}\textcolor{white}{P} &
    \cellcolor{blue}\textcolor{white}{P} &
    \cellcolor{red}\textcolor{white}{L} &
    \cellcolor{red}\textcolor{white}{L} &
    \cellcolor{red}\textcolor{white}{L} &
    \cellcolor{red}\textcolor{white}{L} &
    \cellcolor{red}\textcolor{white}{L} &
    \cellcolor{red}\textcolor{white}{L} \\
    \hline
    \textbf{0.1} &
    \cellcolor{blue}\textcolor{white}{P} &
    \cellcolor{red}\textcolor{white}{L} &
    \cellcolor{red}\textcolor{white}{L} &
    \cellcolor{red}\textcolor{white}{L} &
    \cellcolor{red}\textcolor{white}{L} &
    \cellcolor{red}\textcolor{white}{L} &
    \cellcolor{red}\textcolor{white}{L} &
    \cellcolor{red}\textcolor{white}{L} \\
    \hline
    \multicolumn{1}{|c|}{\diagbox[dir=NE,height=2.2\line]{$q_b$}{$e_b$}} & \textbf{0.0} & \textbf{0.1} & \textbf{0.2} & \textbf{0.3} & \textbf{0.4} & \textbf{0.5} & \textbf{0.6} & \textbf{0.8} \\
    \hline
\end{tabular}
\caption{Whether the CBD about a binary of given $q_b$ and $e_b$ is locked (L; red) or precessing (P; blue), determined from the time-series of the cavity eccentricity vector (\autoref{eqn:ecc_vector}) at $a \leq 5a_b$ (adapted from DeLaurentiis \& Rafikov, in preparation). This locked/precessing partition closely matches the time-stable/time-varying split of $\lambda(t)$ in \autoref{tab:stable_varying_grid}.}
\label{tab:locked_precessing_grid}
\end{table}

In \autoref{tab:stable_varying_grid} we delineate whether $\lambda(t)$ is time-varying or time-stable via a blue cell with a V or red cell with an S, respectively. In \autoref{tab:locked_precessing_grid} we delineate wether the CBD about a given binary is precessing or locked via a blue cell with a P or red cell with an L, respectively. It is of particular note how similar \autoref{tab:stable_varying_grid}, which depicts the time-variability of $\lambda(t)$, is to \autoref{tab:locked_precessing_grid}, which depicts the precession state of the CBD (further discussed in \S\ref{sec:preliminary_analysis} below). For non-circular binaries, time-varying $\lambda(t)$ corresponds to a precessing CBD and time-stable $\lambda(t)$ corresponds to a locked CBD (by ``locked'' we mean a CBD whose orientation, the direction of its eccentric cavity, is fixed relative to the binary, i.e.\ stationary in the binary's co-rotating frame, rather than freely precessing relative to it), suggesting that the time-variability of preferential accretion is, in some part, paced by the precession of the CBD and thereby the cavity. \Stwentythree already noted that forced precession in eccentric binaries is associated with strong modulation of the individual accretion rates on the precession timescale (their Section 3.5), invoking symmetry arguments to argue that circular binaries should remain time-stable while eccentric, forced-precessing binaries should display periodically fluctuating preferential accretion. Our \autoref{tab:stable_varying_grid} extends this picture systematically across the full $(e_b, q_b)$ grid, mapping every simulation in the suite onto the time-stable / time-varying dichotomy and matching it directly to the locked / precessing partition of the CBD.

The non-monotonic dependence of the CBD's precession state on $e_b$ (precessing at low $e_b$, locked at intermediate $e_b$, and precessing again at higher $e_b$) echoes the behavior found by \citet{miranda_munoz_lai_2017} for $q_b = 1$ binaries: their circumbinary discs precess at low and high binary eccentricity but lock to the binary's apsidal line at intermediate eccentricity ($e_b \approx 0.2$--$0.4$). This is consistent with our $q_b = 1$ row, where the CBD is locked at $e_b = 0.2$ yet precessing at $e_b = 0$ and $0.1$ and again at higher eccentricities. \citet{miranda_munoz_lai_2017} describe this intermediate-$e_b$ locking as ``puzzling and unexpected'' and explore several explanations without reaching a firm conclusion: secular (test-particle) theory predicts no apsidal alignment for equal-mass binaries, which lack an octupole potential, while the eccentric Lindblad resonances that would otherwise pump disc eccentricity weaken against viscous damping at high $e_b$. The physical origin of the low-$e_b$ precession-to-locking transition therefore remains an open question.

The $e_b = 0$ simulations, which represent the only differences between \autoref{tab:stable_varying_grid} and \autoref{tab:locked_precessing_grid}, are the natural test of this picture: their CBDs precess freely yet their $\lambda(t)$ is constant. This is consistent with \Stwentythree's symmetry argument: the azimuthal symmetry of a circular binary orbit prevents the precessing CBD from imprinting its variability on the relative accretion rates, even though the disk itself is precessing. It is also consistent with the findings of \citet{DeLaurentiis24}, who studied a different but analogous setup: instead of varying $e_b$ at fixed (non-precessing) binary, they fixed the binary on an eccentric orbit and imposed a general-relativistic (GR) apsidal precession on the binary itself. They found that the GR precession of the binary's pericenter introduces a dominant modulation in the accretion rate, but only when the binary is eccentric enough that the pericenter direction matters; circular binaries cannot translate the precession to preferential accretion at all. Our finding here --- that $e_b = 0$ binaries fail to develop $\lambda(t)$ variability despite a freely precessing CBD --- is the disk-precession analogue of their binary-precession result: in both cases, a non-zero binary eccentricity is required for any precession (of the disk or of the binary) to imprint itself on the relative accretion rates. Equivalently, only the relative position of the binary and the CBD matter for $\lambda(t)$.

Next, we note that the amplitude of the $\lambda(t)$ oscillations are not uniform among the time-varying simulations. In fact, there is a clear correlation between the amplitude of the $\lambda(t)$ oscillation and the $e_b$ of the binary.

\begin{figure}
    \centering
    \includegraphics[width=1\linewidth]{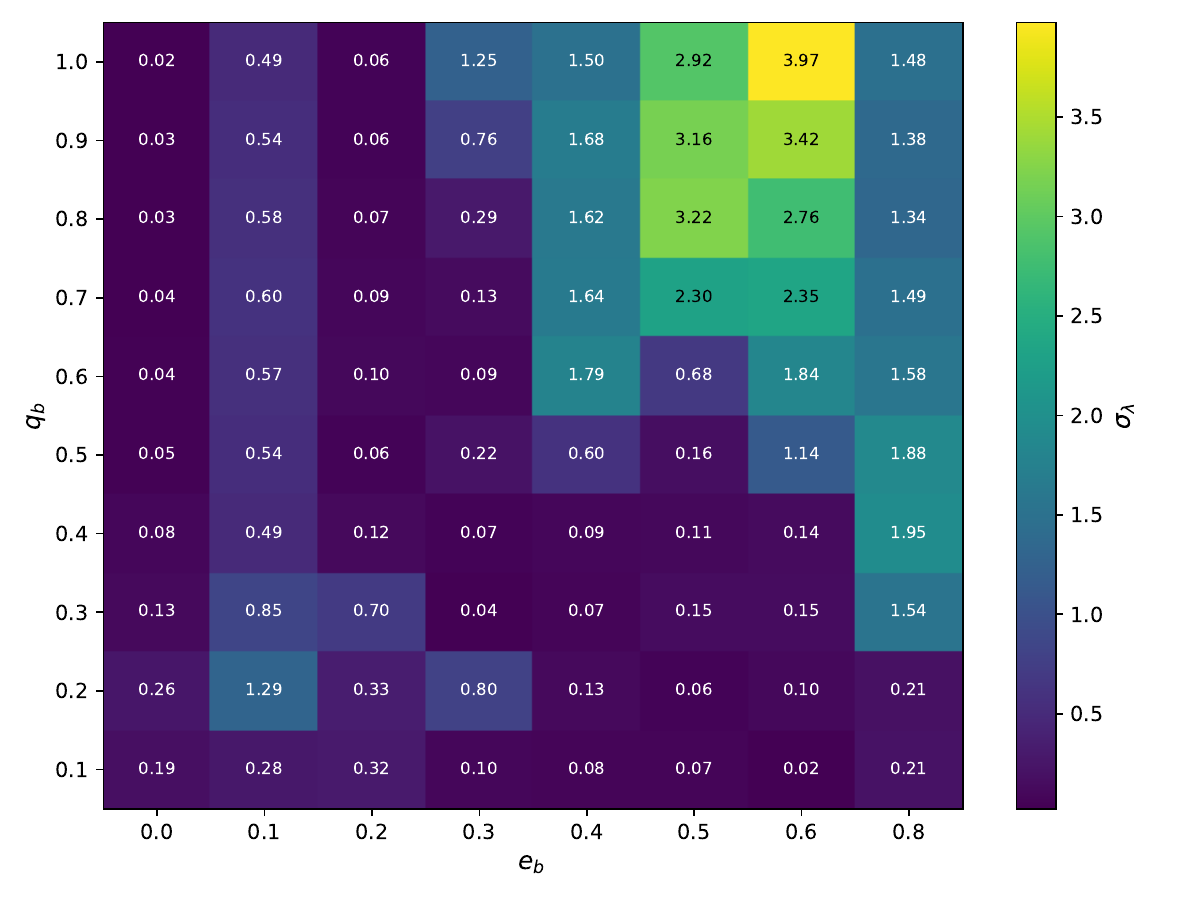}
    \caption{Heat-map of $\sigma_{\lambda}$, the standard deviation of $\lambda(t)$ evaluated over $7000 \tau_b$. Larger values (brighter) indicate stronger time-variability of preferential accretion. The variability peaks broadly in the high-$e_b$, high-$q_b$ region of parameter space.}
    \label{fig:lambda_std_heatmap}
\end{figure}

\begin{figure}
    \centering
    \includegraphics[width=1\columnwidth]{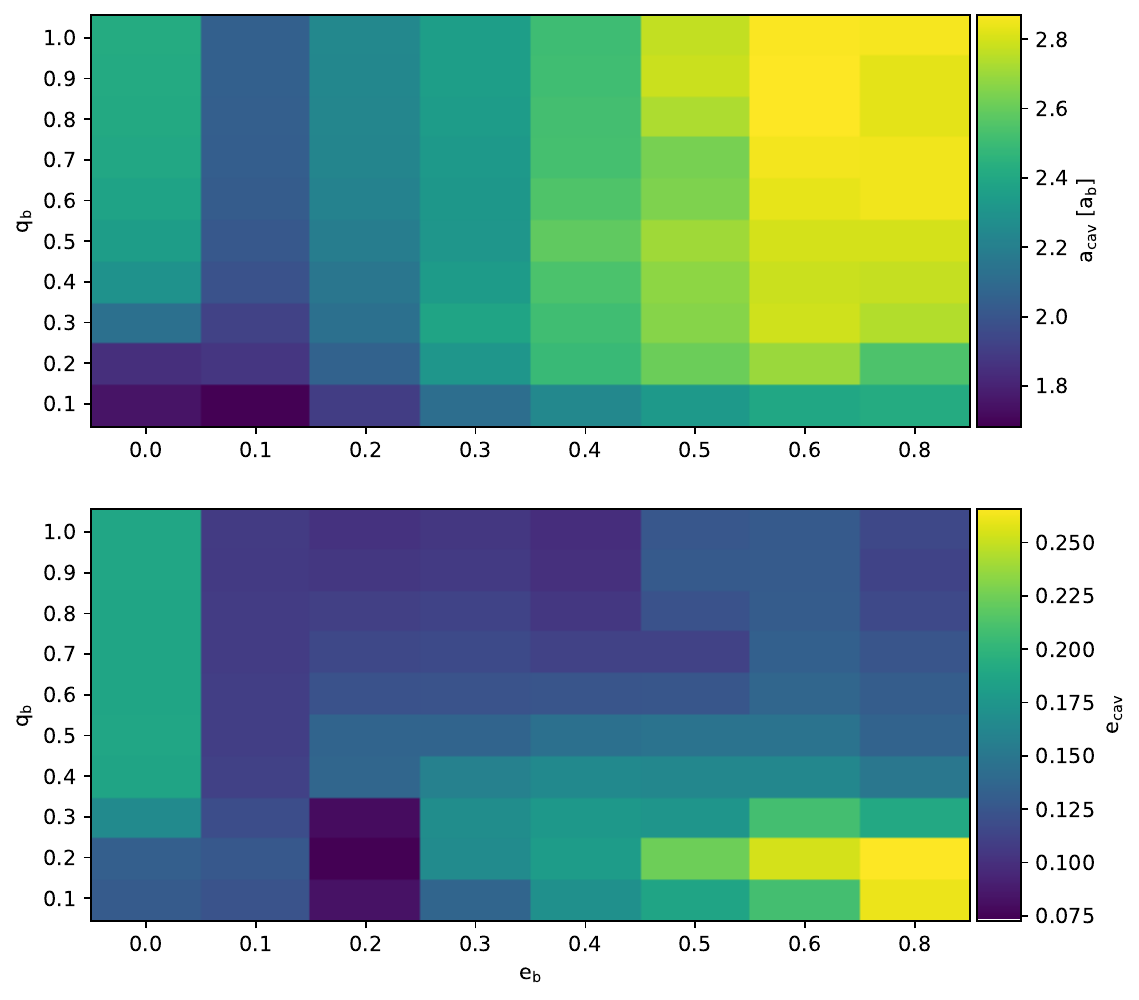}
    \caption{The semi-major axis (top) and eccentricity (bottom) of the cavity for our simulations with varying binary eccentricity and mass ratio, characterized by the $m=1$ Fourier mode of the CBD surface density $\Sigma(r,\theta)$ (adapted from DeLaurentiis \& Rafikov, in preparation). The cavity eccentricity grows with $e_b$, just as $\sigma_\lambda$, as depicted in \autoref{fig:lambda_std_heatmap}.}
    \label{fig:a_e_cav_heatmap}
\end{figure}

In \autoref{fig:lambda_std_heatmap} we display a heat-map indicating the variability amplitude of $\lambda(t)$, quantified by the standard deviation $\sigma_{\lambda}$ over the post-transient window $3000 \le t/\tau_b \le 10\,000$ ($N = 700$ samples at the $10 \tau_b$ snapshot cadence)\footnote{The first $3000 \tau_b$ of each time-series are discarded to remove the initial disk-instability transient documented in DeLaurentiis \& Rafikov (in preparation); the same cut is applied to every $\lambda(t)$ statistic reported in this paper.}. We note that while many of our simulations have small $\sigma_{\lambda}$ since they are time-stable (see \autoref{tab:stable_varying_grid}), those that display meaningful $\sigma_{\lambda}$ suggest a trend. Namely, we find that $\sigma_{\lambda}$ increases with $e_b$, peaking at $e_b = 0.6$. This modest increase in $\sigma_{\lambda}$ with $e_b$ mirrors a corresponding trend in the cavity eccentricity. \autoref{fig:a_e_cav_heatmap} shows the cavity semi-major axis and eccentricity across the suite: the cavity eccentricity grows with $e_b$ and peaks near $e_b = 0.6$, closely tracking $\sigma_\lambda$. This correlation further suggests that the CBD is an important regulator of preferential accretion.

\begin{figure}
    \centering
    \includegraphics[width=1\linewidth]{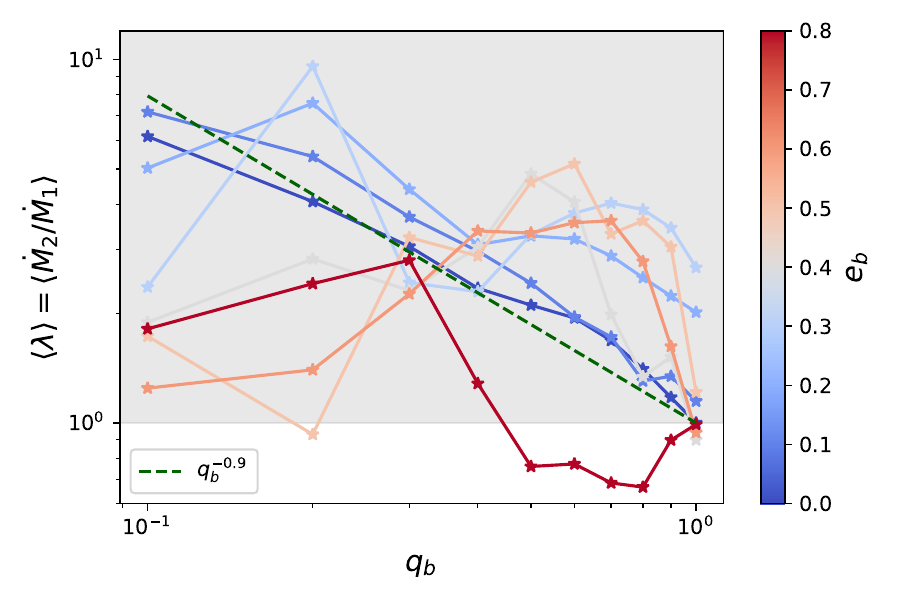}
    \caption{Time-averaged accretion-rate ratio $\langle \lambda \rangle = \langle \dot{M}_2 / \dot{M}_1 \rangle$ as a function of $q_b$ (x-axis) and $e_b$ (line color). The gray-shaded region marks $\langle\lambda\rangle>1$, where the simulation labelled ``secondary'' accretes preferentially. The dashed green line is the $q_b^{-0.9}$ power-law fit reported by \Stwentythree for circular binaries. Preferential accretion is always onto the secondary, with low-$e_b$ binaries following the $q_b^{-0.9}$ trend.}
    \label{fig:mean_lambda_lines}
\end{figure}

\begin{figure}
    \centering
    \includegraphics[width=1\linewidth]{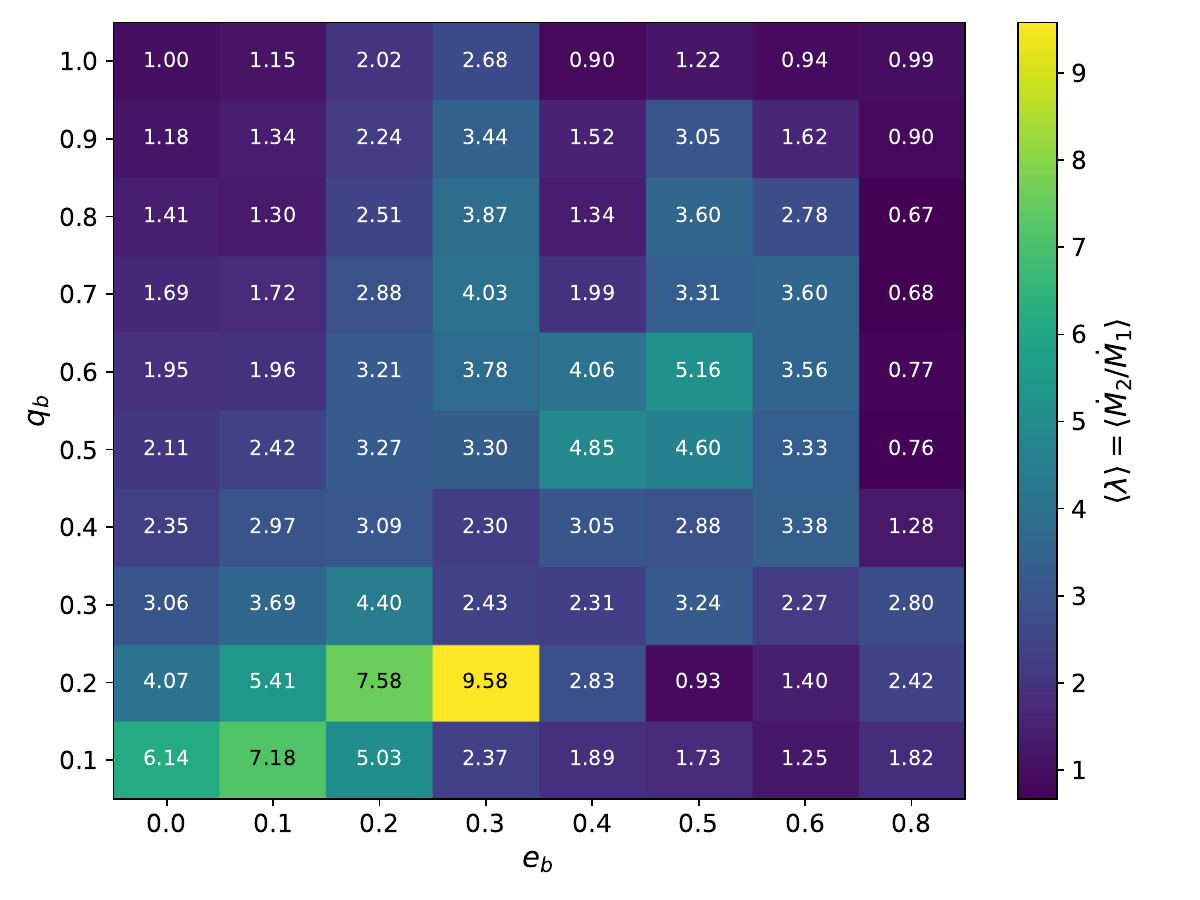}
    \caption{Heat-map of the time-averaged accretion-rate ratio $\langle\lambda\rangle = \langle\dot{M}_2/\dot{M}_1\rangle$ across the simulation suite. The x-axis is the binary eccentricity $e_b$ and the y-axis is the binary mass-ratio $q_b$; in-cell labels show $\langle\lambda\rangle$. Preferential accretion is strongest for low-$e_b$, low-$q_b$ binaries and weakest for high-$e_b$, high-$q_b$ binaries.}
    \label{fig:lambda_mean_heatmap}
\end{figure}

In addition to variability of $\lambda(t)$ we also comment on its mean value, determining which BH is preferred to accrete and to what extent. We report our results in \autoref{fig:mean_lambda_lines} and \autoref{fig:lambda_mean_heatmap}.

In \autoref{fig:mean_lambda_lines} we display the mean values of $\lambda(t)$ against $q_b$, coloured by $e_b$. We find that our figure is in broad agreement with Figure 4 of \Stwentythree. In particular, $\langle \lambda \rangle$ of the $e_b=0$ simulation approximately follows the power law of $q_b^{-0.9}$. However, we also find that $e_b=0.1$ also holds true to this line, suggesting that the behavior of $\langle \lambda \rangle $ is generic to lower eccentricity binaries and not unique to $e_b=0$. At higher eccentricity, however, this monotonic $q_b^{-0.9}$ trend breaks down: the $e_b \gtrsim 0.5$ curves are non-monotonic in $q_b$, rising to a peak at intermediate $q_b$ before falling. Further, we find that low and intermediate $e_b$ binaries have $\lambda >1$, while high-$e_b$, high $q_b$ binaries have $\lambda <1$. This is in line with all the results of the literature \citep{munoz_miranda_lai_19, duffell_dorazio_2020, farris_2014, Dittmann_Ryan_21, siwek_prefacc}. We note that the extent to which one BH accretes over the other depends greatly on the binary parameters, as will be discussed further in \autoref{sec:preliminary_analysis}, below.

The dependence of $\langle \lambda \rangle$ on $q_b$ and $e_b$ is shown in 2D in \autoref{fig:lambda_mean_heatmap}, which displays $q_b$ on the y-axis and $e_b$ on the x-axis, encoding the magnitude of preferential accretion by color. At first glance it is clear that some binary parameters lend themselves to stronger preferential accretion than others: low-$e_b$, low-$q_b$ binaries display the largest values, while high-$e_b$, high-$q_b$ binaries accrete more equally, in agreement with \Stwentythree. This is opposite to $\sigma_{\lambda}$ values in \autoref{fig:lambda_std_heatmap}, which increases with $e_b$ and $q_b$ and the eccentricity of the CBD in \autoref{fig:a_e_cav_heatmap} which increases with $e_b$ and decreases with $q_b$. This suggests that the binary parameters themselves are more important than the CBD in regulating the level of preferential accretion.

While there is a trend, the level of preferential accretion is by no means monotonic with respect to $q_b$ and $e_b$. We note that intermediate values (e.g.\ $(e_b, q_b) = (0.4, 0.4)$ and $(0.4, 0.5)$) have higher preferential accretion levels than their neighbors. Further, \autoref{fig:lambda_mean_heatmap} displays interesting ``hotspots'' of high preferential accretion, with $(e_b, q_b) = (0.3, 0.2)$ being the simulation with the highest levels of preferential accretion. We also find curious dimspots where the accretion onto both BHs is near-equal despite the binary having unequal mass. Notably, $(e_b, q_b) = (0.5, 0.2)$ has $\langle \lambda \rangle = 0.93$. We emphasize that cells with $\langle\lambda\rangle$ slightly \emph{below} unity should not be read as evidence that the primary out-accretes the secondary. For these cells the $\lambda(t)$ time-series in \autoref{fig:lambda_full} oscillates about unity without a sustained preference for either BH; whether the time-average lands just above or just below unity is then set by the (arbitrary) apocenter labelling of which BH is the ``primary,'' so $\langle\lambda\rangle$ marginally below unity does not indicate genuine primary preference. Cells with $\langle\lambda\rangle > 1$, by contrast, signal genuine preferential accretion, and even small offsets above unity can be physically significant: as we discuss in \S\ref{sec:mass_ratio}, the $q_b = 1$ simulations at $e_b = 0.2$ and $0.3$ have $\langle\lambda\rangle$ only modestly above$\ 1$ yet still drive the binary away from equal mass.

More importantly, we note a point of deviation from \Stwentythree with direct consequences for the binary's mass-ratio evolution. For the $(e_b, q_b) = (0.2, 1.0)$ and $(0.3, 1.0)$ simulations we find $\langle \lambda \rangle \neq 1$ (\autoref{fig:mean_lambda_lines}, \autoref{fig:lambda_mean_heatmap}): nominally equal-mass binaries that nonetheless accrete preferentially onto one BH and therefore evolve away from $q_b = 1$. The $(0.2, 1.0)$ case was reported by \Stwentythree to remain at $q_b = 1$; that simulation contained a numerical error, and once it is corrected we recover preferential accretion. We note that this numerical error is limited in scope and does not affect any other results in \Stwentythree and related papers. We discuss the resulting drift away from equal mass (and its implications for the SMBBH mass-ratio distribution) in \autoref{sec:mass_ratio} and \autoref{sec:observations}.

Clearly, there is variability in both the fluctuations of preferential accretion and its mean value across simulations. In the following section we make a preliminary attempt to understand preferential accretion as an effect of the unique geometry between the binary and the CBD.

\subsubsection{Physical origin of preferential accretion}\label{sec:preliminary_analysis}

In \autoref{subsec:prefer-describe} we noted similarities between the preferential accretion behavior $\lambda(t)$ and the CBD geometry. The delineation of time-stable and time-varying $\lambda(t)$ in \autoref{tab:stable_varying_grid} is similar to that between precessing and locked circumbinary disks. While alternative explanations have been proposed \citep{Artimowicz_83_og_polish}, the assumption that the preferential accretion of the binary is related to the relative closeness of the components to the CBD's inner edge \citep{dorazio_2013,Rafikov_16_accretion, farris_2014} has been left unchallenged. In the following section we provide a first, preliminary, analysis of this hypothesis and alternative ways in which the CBD could influence preferential accretion.

\begin{figure}
    \centering
    \begin{tikzpicture}[scale=2.0]
      \draw[thick, black] (0,0) ellipse (1.4 and 1.0);

      \node[black, anchor=center, font=\scriptsize\itshape] at (0, 0.78) {CBD inner edge};

      \draw[gray, dashed, thin] (-0.109, 0) -- (+1.091, 0);

      \draw[<->, blue!70!black, thick] (-0.109, 0) -- (-1.4, 0);
      \draw[<->, red!70!black, thick]  (+1.091, 0) -- (+1.4, 0);

      \draw[black, thick] (-0.04, -0.04) -- (0.04, 0.04);
      \draw[black, thick] (-0.04, 0.04) -- (0.04, -0.04);

      \fill[blue!70!black] (-0.109, 0) circle (0.06);
      \fill[red!70!black]  (+1.091, 0) circle (0.030);

      \node[blue!70!black, anchor=south, font=\scriptsize] at (-0.109, 0.07) {$M_1$};
      \node[red!70!black,  anchor=south, font=\scriptsize] at (+1.091, 0.05) {$M_2$};
      \node[black, anchor=north, font=\scriptsize] at (0, -0.10) {COM};
      \node[blue!70!black, anchor=north, font=\scriptsize] at (-0.75, -0.03) {$r_1$};
      \node[red!70!black,  anchor=north, font=\scriptsize] at (+1.25, -0.03) {$r_2$};

      \draw[->, gray!60!black, thick] (-0.50, 1.10) arc[start angle=110, end angle=70, radius=1.5];
      \node[gray!60!black, anchor=south, font=\scriptsize\itshape] at (0, 1.20) {precession};
    \end{tikzpicture}
    \caption{Schematic illustration of the cavity geometry with the binary at apocenter for the precessing CBD case, illustrated for an unequal-mass binary with $q_b \ll 1$. The CBD inner edge (black ellipse) is closer to the secondary BH ($M_2$, red) on one side; the primary ($M_1$, blue) is farther from its nearby cavity wall. Closest distances from each BH to the wall are $r_1$ and $r_2$, with $r_1 > r_2$ in the locked configuration shown. As the disk precesses in the binary's frame (curved gray arrow above the ellipse), the ellipse's major axis rotates, and the vectors $\vec{r}_1$ and $\vec{r}_2$ change accordingly.}
    \label{fig:cavity_cartoon}
\end{figure}
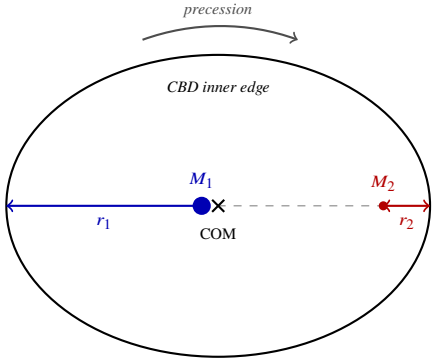

As discussed in \autoref{sec:methods}, we use the dominant Fourier modes of the cavity (specifically the $m=1$ mode of the surface density $\Sigma(r,\theta)$ at its inner edge) to reconstruct its shape as a function of azimuthal angle $\theta$: $r_{\rm cav}(\theta)$, with the origin at the binary center of mass. Combined with the position of each BH at the apocenter snapshot, we define the vector from each BH to the nearest point of the cavity inner edge. Taking $\theta_1^*$ and $\theta_2^*$ to be the angles that minimize the distance from each BH to the cavity wall, we write $\vec{r}_1 = \vec{r}_{\rm cav}(\theta_1^*) - \vec{r}_{\rm BH_1}$ and $\vec{r}_2 = \vec{r}_{\rm cav}(\theta_2^*) - \vec{r}_{\rm BH_2}$ for the primary and secondary respectively, with magnitudes $r_1 = |\vec{r}_1|$ and $r_2 = |\vec{r}_2|$ (see \autoref{fig:cavity_cartoon}). We emphasise that our snapshots are, unfortunately, taken only at apocenter; the vectors $\vec{r}_1$ and $\vec{r}_2$ are not literal closest-approach distances over the full orbit but a single-phase snapshot of the cavity orientation relative to the binary axis. As the disk precesses, the orientation of $r_{\rm cav}(\theta)$ relative to the fixed apocenter line changes, and $r_1(t)$ and $r_2(t)$ inherit that variability. They therefore serve as a proxy for cavity orientation, not as a moment-by-moment proximity metric.

\begin{figure}
    \centering
    \includegraphics[width=1\linewidth]{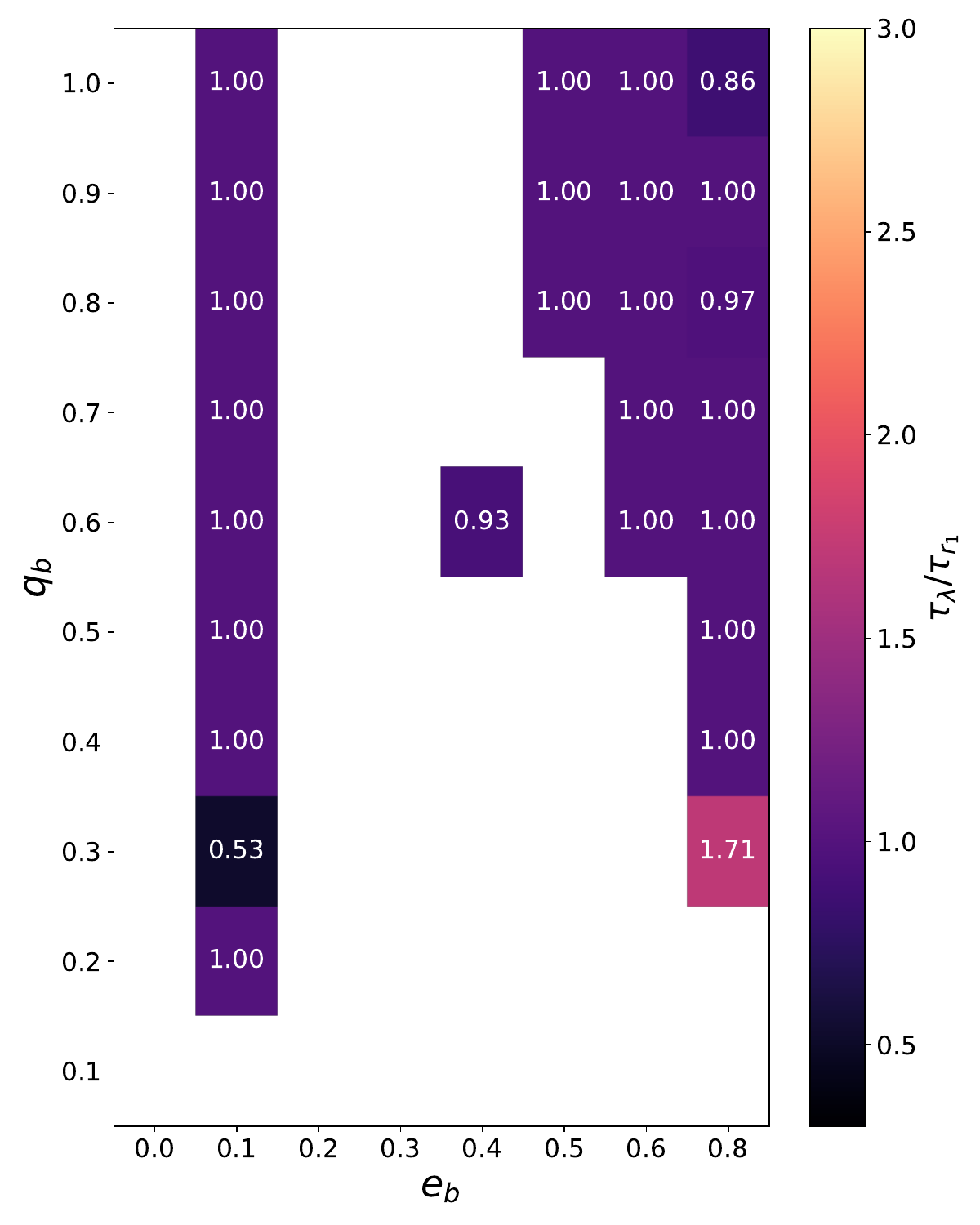}
    \caption{The ratio of the period of preferential-accretion variability ($\tau_\lambda$) to the period of the CBD precession ($\tau_{r_1}$). The x-axis is the binary eccentricity, the y-axis the binary mass-ratio, and the cell colour the ratio of the two periods. Uncoloured cells are those where no clear period could be determined for $r_1$ or $\lambda$: either because the signal is time-stable (a locked CBD) or because it is too irregular for a single dominant Fourier peak to be believably identified. Where both periods are defined they are nearly always equal (ratio ${\approx}1$), tying the $\lambda(t)$ variability to the cavity precession.}
    \label{fig:rmin_lambda_peak_ratio}
\end{figure}

A first step in determining the relationship between preferential accretion and the proximity to the CBD is to study the temporal behavior of these quantities. We have already determined that, except for the $e_b = 0$ case, all precessing CBDs result in a time-varying $\lambda(t)$. $r_{1}(t)$ and $r_{2}(t)$, proxies for the CBD orientation, display the same split into time-stable and time-varying behavior as $\lambda(t)$ in \autoref{tab:stable_varying_grid}. For simulations with time-varying $\lambda(t)$ we compare their period of oscillation with that of $r_{1}$ and $r_{2}$. To determine the period of each time-series we first calculate a fast Fourier transform (FFT) and normalize the amplitudes by their sum. The period is defined as that of the largest-amplitude peak in the spectrum exceeding a normalized value of $0.05$. We find the periods of $r_1$\footnote{It is important to note that we have found the period of $r_1$ and $r_2$ to be exactly equal in all our simulations, making either appropriate for this calculation.} and $\lambda$ in this fashion and display the ratio of the two periods in \autoref{fig:rmin_lambda_peak_ratio}. This result is insensitive to the precise amplitude cut: adopting $0.03$ or $0.07$ in place of $0.05$ changes only how many cells yield a well-defined period (more at the lower cut, fewer at the higher), while in every cell where both periods are defined the ratio remains $\approx 1$.

The ratios reported in \autoref{fig:rmin_lambda_peak_ratio} are nearly all equal to unity. Though certain binaries have ratios that deviate from unity, these deviations do not suggest resonances between CBD precession and preferential accretion but rather point to the FFT being unstable when applied to periodic, non-sinusoidal $\lambda(t)$ (e.g.\ $(e_b, q_b) = (0.4, 0.6)$). Thus, the precession of the CBD and the variability of $\lambda(t)$ are tightly correlated in both occurrence and period: we observe the same FFT period in both quantities and the same locked-vs-precessing partition.

\begin{figure*}
    \centering
    \includegraphics[width=1\linewidth]{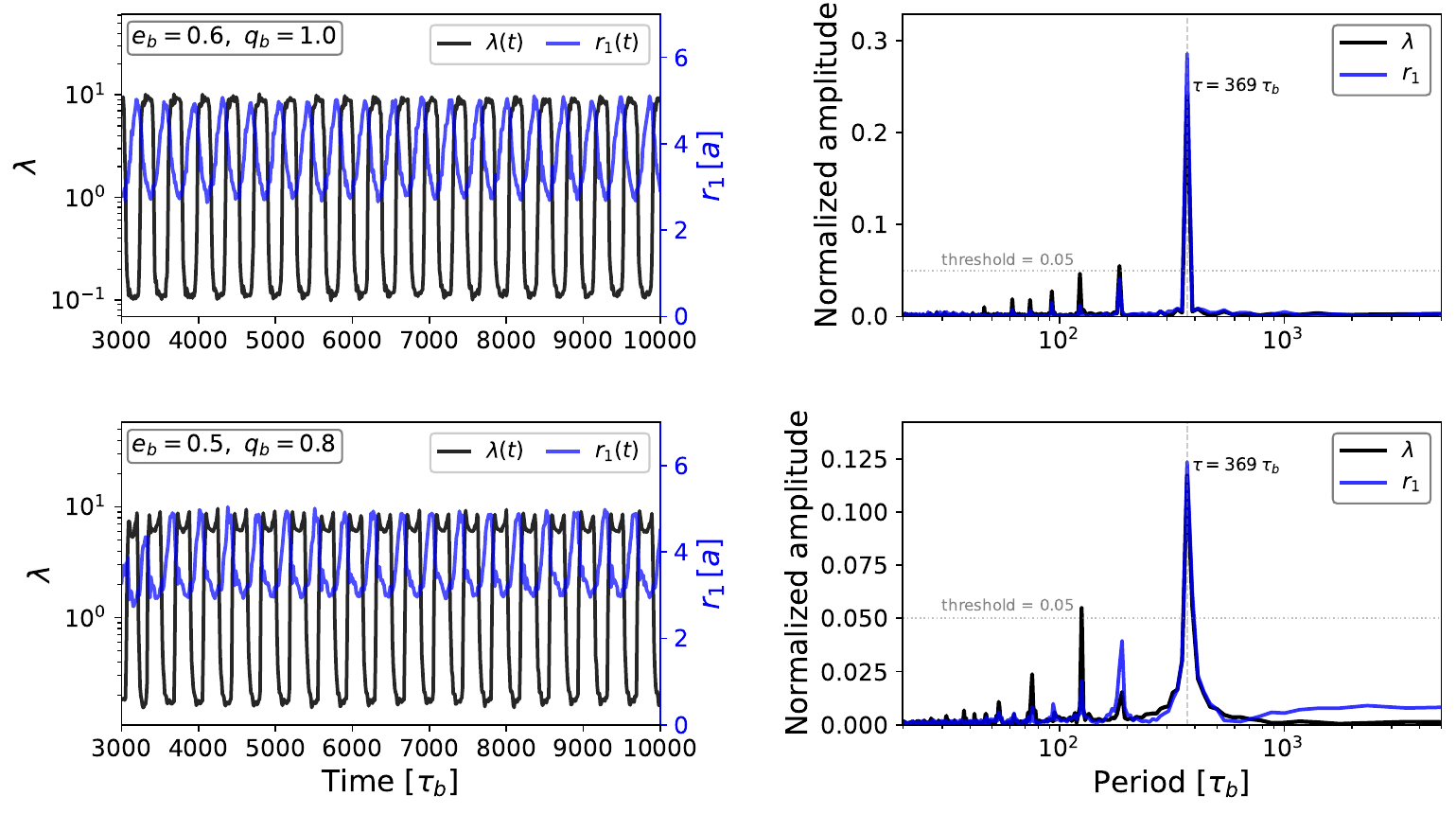}
    \caption{Time-series and power spectra for $(e_b, q_b) = (0.6, 1.0)$ (top row) and $(0.5, 0.8)$ (bottom row), illustrating the period match between $\lambda(t)$ and the cavity-orientation proxy $r_1(t)$. Left column: $\lambda(t)$ (black, left log-axis) and $r_1(t)$ (blue, right axis in units of $a$) over the post-transient window $3000 \le t/\tau_b \le 10000$ ($N=700$ samples at the snapshot cadence of $10\tau_b$). Right column: normalized FFT amplitudes $|\tilde X(f)| / \sum_f |\tilde X(f)|$ as a function of period $\tau = 1/f$; the dotted line marks the $0.05$ amplitude threshold used in \autoref{fig:rmin_lambda_peak_ratio}. Both signals peak cleanly at $\tau \approx 369 \tau_b$ in each simulation (marked by the vertical dashed line), demonstrating that $\lambda(t)$ variability tracks the precessing cavity.}
    \label{fig:fourier_panels}
\end{figure*}

Because our snapshots are recorded only at apocenter, the time-series $r_1(t)$ and $r_2(t)$ inherit the cavity's precession frequency: each snapshot catches the cavity at a slightly rotated orientation. \autoref{fig:fourier_panels} illustrates the resulting period match for two representative simulations, $(e_b, q_b) = (0.6, 1.0)$ and $(0.5, 0.8)$. In both cases, $\lambda(t)$ and $r_1(t)$ peak at the same period $\tau \approx 369 \tau_b$ above the $0.05$ amplitude threshold. This match justifies the inference that $\lambda(t)$ variability is paced by the cavity's apsidal precession.

\begin{figure}
    \centering
    \includegraphics[width=1\linewidth]{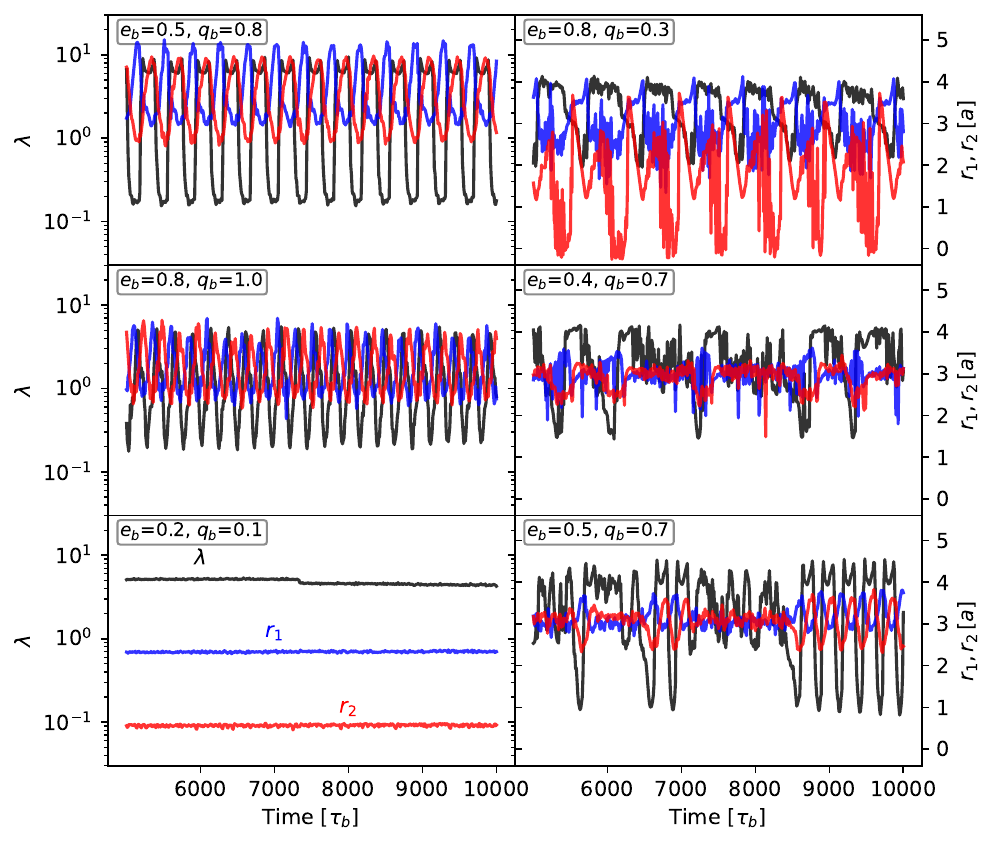}
    \caption{Comparison between the time-series of $\lambda(t)$ (black, left log-axis), $r_1(t)$ (blue), and $r_2(t)$ (red) (right axis, in units of the binary semi-major axis $a$), where $r_1$ and $r_2$ are the shortest distances from the primary and secondary, respectively, to the cavity inner edge at the apocenter snapshot (a proxy for cavity orientation; see \S\ref{sec:preliminary_analysis}). Time is in binary orbital periods $\tau_b$; panels are plotted for $5000 \le t/\tau_b \le 10000$. Clockwise from upper-left: $(e_b, q_b) = (0.5, 0.8)$, $(0.8, 0.3)$, $(0.4, 0.7)$, $(0.5, 0.7)$, $(0.2, 0.1)$, and $(0.8, 1.0)$. We find that the BHs' position with respect to the cavity does not directly determine the components' relative accretion rates.}
    \label{fig:lambda_rmin_panels}
\end{figure}

In \autoref{fig:lambda_rmin_panels} we display $r_1$ (blue) $r_2$ (red) and $\lambda$ (black) for a slice in time for six of our simulations, with the dual y-axis representing $\lambda$ (left) and the distance from each BH to the edge of the cavity in units of $a_b$ (right). As indicated by \autoref{fig:rmin_lambda_peak_ratio}, the variability of the distance from the BHs to the cavity displays the same period (or lack thereof) as $\lambda$. This is clear for the simulations in the left column of \autoref{fig:lambda_rmin_panels}. However, we note that for some of the simulations in the right column the period is unstable. This behavior is especially well captured by $(e_b, q_b) = (0.5, 0.7)$, where the change from a stochastic to periodic $\lambda$ signal at $\approx 8500 \tau_b$ is mirrored in the change of behavior of $r_1$ and $r_2$. Further, in $(e_b, q_b) = (0.4, 0.7)$ we find that large amplitude changes in $r_1$ and $r_2$ (e.g.\ at $t \approx 6000$ and $9000 \tau_b$) occur simultaneously with large amplitude changes in $\lambda$.

The naive understanding of the tie between preferential accretion and the distance to the black holes is as follows. The inner cavity rim hosts the densest gas in the system, and the relative velocity between a BH and the rim gas is lowest when the BH is closest to the rim \citep{shi_2012, farris_2014, duffell_dorazio_2020}. Whenever a black hole's Hill sphere overlaps that rim, the steep local gravitational-potential gradient bends incoming streamlines into its mini-disk, and the instantaneous capture rate obeys a Bondi-Hoyle-like scaling
\begin{equation}
  \dot{M}_{\rm near} \propto \Sigma_{\rm rim} \, v_{\rm rel}^{-3}.
\end{equation}
A companion farther from the rim encounters lower surface density and higher gas-BH relative velocity \citep{miranda_munoz_lai_2017, dorazio_duffel}, so
\begin{equation}
  \frac{\dot{M}_{\rm far}}{\dot{M}_{\rm near}} \sim \frac{\Sigma}{\Sigma_{\rm rim}} \left(\frac{v_{\rm rel}}{v_{\rm rel,\,near}}\right)^{3} \ll 1,
\end{equation}
suppressing its accretion by orders of magnitude even though both black holes share the same global gas reservoir.

The $(e_b, q_b) = (0.2, 0.1)$ case illustrate what we would expect from this naive explanation. The corresponding panel of \autoref{fig:lambda_rmin_panels} shows that this simulation has a CBD that is locked such that the primary black hole (blue) is about three times as far from the cavity edge than the secondary black hole (red). Correspondingly, we see that the secondary is accreting at a rate about five times the primary, seemingly as a result of the secondary's greater pull on its neighboring portion of the CBD. In fact, we can even take this further and note that the tidal field of the secondary on its nearby cavity wall is approximately four to five times stronger than the tidal field of the primary on its more distant cavity wall (with the tidal field scaling as $M/r^3$ and $r_1/r_2 \approx 3.4$ measured from the simulation snapshot), in suggestive agreement with the observed factor of $\sim 5$ in the relative accretion rate. However, the naive approximation is inherently over-simplistic. It does not take into account each BH's minidisk, the streams by which the gas is fed to the BH, or any non-linear fluid dynamics, such as shocks, that may be relevant to the feeding of each BH's minidisk. Studying the other panels, indeed we see that the naive explanation no longer suffices.

Before discussing the deviations, we make the naive prediction explicit. If preferential accretion were governed solely by proximity to the cavity wall, then for a binary at apocenter with the cavity oriented toward the secondary, $r_2$ should be at its minimum (cavity wall closest to secondary) precisely when $\lambda$ is at its maximum (secondary accreting most). As the cavity precesses, $r_2$ should rise to its maximum a half-precession-period later, when $\lambda$ should be at its minimum. We therefore expect $r_2(t)$ to be \textit{exactly $\pi$ out of phase} with $\lambda(t)$, while $r_1(t)$ (by symmetry, since the cavity wall is then closest to the primary) should be \textit{exactly in phase} with $\lambda(t)$.

All simulations displayed in \autoref{fig:lambda_rmin_panels}, except for $(e_b, q_b) = (0.2, 0.1)$, show that lower $r_2$ values do not correspond to higher accretion rates onto the secondary. The simulations $(e_b, q_b) = (0.5, 0.8)$ and $(0.8, 1.0)$ clearly do not display the above naive behavior: the former shows both $r_1$ and $r_2$ at $\pi/4$ out of phase with $\lambda$, while the latter shows the exact opposite of the naive expectation, with $r_1$ in phase and $r_2$ out of phase with $\lambda$. Further, $(e_b, q_b) = (0.8, 0.3)$ and $(0.4, 0.7)$ break the naive picture in another way: the amplitude of $r_1$ relative to $r_2$ does not strictly coincide with which BH accretes preferentially. For $(0.8, 0.3)$, while the primary is always at least as far from the cavity wall as the secondary, we still see periods where the primary accretes preferentially. For $(0.4, 0.7)$, during periods where both BHs are equidistant from the cavity wall, the secondary accretes up to 4 times more than the primary\footnote{While $\lambda$ is a ratio and this behavior could be due to small amplitude deviations in $\dot{M}_1$ and $\dot{M}_2$, we found that the amplitude of the sum is not strongly suppressed during this period.}.

From the above analysis it is clear that the simple, instantaneous tidal/proximity picture (in which $r_2(t)$ would be exactly $\pi$ out of phase with $\lambda(t)$) does not hold in detail across the suite. To quantify the relationship more directly, we compute the normalized, lagged cross-correlation between $\lambda(t)$ and the secondary's cavity-wall distance $r_2(t)$,
\begin{equation}\label{eqn:crosscorr}
    C_{\lambda, r_2}(\Delta t) = \frac{\big\langle \, (\lambda(t) - \langle\lambda\rangle)\,(r_2(t+\Delta t) - \langle r_2\rangle) \, \big\rangle_t}{\sigma_\lambda \, \sigma_{r_2}},
\end{equation}
and report its principal peak, $\max_{\Delta t} C_{\lambda, r_2}$, together with the lag $\Delta t$ at which it occurs (we focus on $r_2$ since, as noted above, $r_1$ and $r_2$ are tightly anti-correlated proxies for the same cavity orientation, so $r_1$ adds no independent information). For the precessing, time-varying cells the correlation is strong: $18$ of the $80$ simulations reach $\max_{\Delta t} C_{\lambda, r_2} \ge 0.7$ (median $0.88$), with the peak occurring at a small positive lag (median $\Delta t \approx 60\,\tau_b$, or $\approx 50^\circ$ of a precession cycle). The strength of this correlation indicates that in these cells the cavity-wall distance accounts for most of the modulation of preferential accretion; the finite lag shows that $\lambda$ does not respond instantaneously to the cavity geometry, but rather with a delay plausibly set by the time for the accretion flow to react and reach the minidisks/sinks. In the locked or otherwise irregular cells the correlation is weak ($\max_{\Delta t} C_{\lambda, r_2} \lesssim 0.3$), indicating that additional physics of comparable importance operates there. We present the cross-correlation analysis in full in Appendix~\ref{sec:appendix_crosscorr} (\autoref{fig:cc_examples}, \autoref{fig:cc_heatmap}). A more complete explanation of preferential accretion will be developed in a forthcoming paper.

\subsection{Evolution of the mass ratio}\label{sec:mass_ratio}
\begin{figure}
    \centering
    \includegraphics[width=1\linewidth]{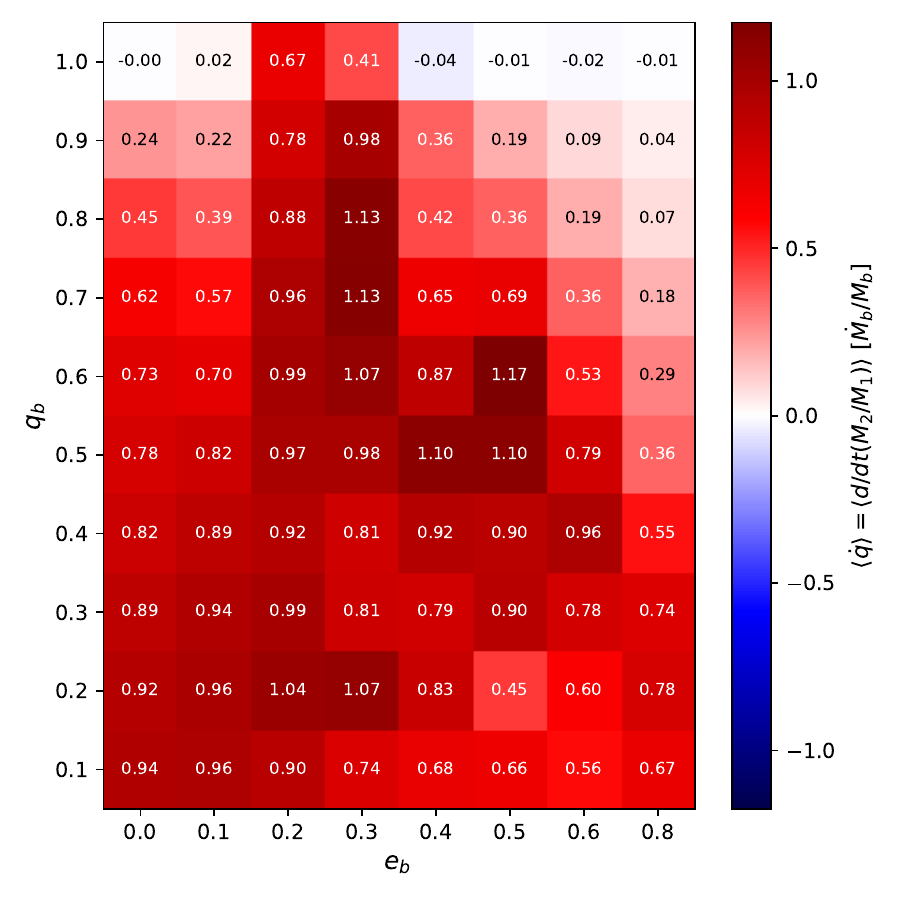}
    \caption{The time-averaged rate of change of the binary mass-ratio, $\langle \dot{q}_b \rangle$, for the simulation suite, in $\dot{M}_b / M_b$ units. The x-axis is the binary eccentricity $e_b$ and the y-axis is the binary mass-ratio $q_b$. The diverging colormap is centered at $\langle \dot{q}_b \rangle = 0$: red cells indicate the binary evolves toward $q = 1$ (positive $\langle \dot{q}_b \rangle$), blue cells away from $q = 1$. Most $q_b < 1$ cells are positive (driven toward equal mass); the $q_b = 1$ row is mostly consistent with zero, with notable exceptions at $e_b = 0.2$ and $0.3$ where the binary evolves away from unity (see \S\ref{sec:mass_ratio} for discussion).}
    \label{fig:qdot_heatmap}
\end{figure}

In addition to preferential accretion, we report the average rate of change of the mass ratio, $\langle \dot{q}_b \rangle$, the time-average of $\dot{q}_b(t)$ over the post-transient portion of each simulation. Namely, we calculate $\dot{q}_b(t)$ as described in \autoref{eqn:q_dot} in \autoref{sec:methods}, make a time-cut at $3000\,\tau_b$ to ensure that early numerical instabilities do not affect our results, and report the mean on the truncated time-series in units of $\dot{M}_b/M_b$ in \autoref{fig:qdot_heatmap}.

A particularly important feature of Fig. 11 is the slow mass-ratio evolution at high eccentricity and large, but sub-unity, mass ratio. For $e_b=0.6$–$0.8$ and $q_b=0.8$–$0.9$, the measured rates imply that changing the mass ratio by approximately $0.1$ would take roughly $20$–$400\,\mathrm{Myr}$ for $\dot{M}=0.3$–$1 \, \dot {M}_{\mathrm{Edd}}$. These binaries still evolve toward equal mass, but they need not reach $q_b=1$ during a finite quasar episode.

This result complements \citet{valli2024}, who showed that substantial evolution in binary separation or mass ratio generally requires the binary to accrete a significant fraction of its initial mass. It is also consistent with \citet{xu2026}, who found that binaries above the low-$q_b$ equilibrium evolve gradually toward equal mass. Their calculation assumes circular binaries and does not follow the coupled evolution of $a_b$, $e_b$, and $q_b$ into the LISA band. We perform that calculation in \autoref{sec:unequal_mass}.

The upper row of \autoref{fig:qdot_heatmap} ($q_b = 1$) requires care, both in labelling and in interpretation. Our convention $q_b \equiv M_2/M_1 \leq 1$ assigns ``primary'' to the more massive BH and ``secondary'' to the less massive one; in a strictly $q_b = 1$ system this assignment is degenerate, and we follow \Stwentythree in identifying the components by their spatial location at apocenter. We retain these original labels throughout the gas-driven phase, even after $q_b = 1$ is broken by accretion, so that ``evolves away from unity'' should be read as: the mass ratio $M_2/M_1$ defined by the original apocenter assignment drifts away from $1$. With this convention, most $q_b = 1$ simulations display $\langle \dot{q}_b \rangle$ values consistent with zero\footnote{The slightly negative values are within the standard error of the mean for the $\dot{q}_b$ time-series ($\sigma \approx 10^{-2}$ for $q_b = 1$, estimated from the variance of $\dot{q}_b(t)$ divided by the integration duration), and can thus be taken to be zero.} The binary remains at equal mass to within statistical noise. Two simulations stand out: $(e_b, q_b) = (0.2, 1.0)$ and $(0.3, 1.0)$ both show strong positive $\langle \dot{q}_b \rangle$, meaning the BH initially labeled ``secondary'' grows into the more massive component, and the binary is evolving away from unity. This stands as a correction to \Stwentythree, where the $(0.2, 1.0)$ case was reported to remain at $q_b = 1$; after ensuring correct sink-particle tracking in the $q_b = 1$ simulations, we find that accretion toward equal mass is not a foregone conclusion. The result is consistent with DeLaurentiis \& Rafikov (in preparation), who report that the $(0.2, 1.0)$ CBD is stably locked (as in \Stwentythree) with a non-varying $\lambda$ and the primary further from the cavity edge than the secondary.

The finding that equal-mass binaries at $e_b = 0.2, \, 0.3$ accrete away from equal mass has implications for CBD structure and SMBBH population statistics. We note that both the $(e_b, q_b) = (0.2, 1.0)$ and the $(0.2, 0.9)$ simulations have locked disks in roughly the same orientation, with the pericenter of the disk closest to the secondary. As the $q_b = 1$ case accretes away from unity, the BH initially identified as the ``secondary'' grows into the primary; the disk, oriented toward the original secondary, must therefore realign itself, flipping in concert with the switch in primary and secondary identities. We speculate that this realignment proceeds on the disk's apsidal precession timescale, since the same precession dynamics that orient locked disks in the first place are the natural mechanism by which a locked disk can re-orient. This precession timescale is plausibly much shorter than the AGN-disk lifetime; if so, the cavity re-orients rapidly compared with the gas-driven mass-ratio evolution, which would keep the lookup of \autoref{fig:qdot_heatmap} applicable and hold the binary close to (though not exactly at) $q = 1$. The details of this reaction would provide insight into the CBD-orientation mechanism, into how far from unity the binary ultimately evolves, and, depending on the geometry and timescale of re-orientation, could constitute an event with characteristic EM signatures. Confirming this picture would require live-binary simulations through a sustained $\dot{q}_b \neq 0$ phase, which we leave to future work.

\section{Observational implications}\label{sec:observations}
In the following section we discuss the potential observational consequences of our $\lambda(t)$ and $\dot{q}_b$ results (see \autoref{sec:results}).

\subsection{Jet launching}

A key observational consequence of accretion onto BHs is the possible
launching of relativistic jets. Jet launching is fundamentally magnetic,
rather than a direct consequence of radiative inefficiency. In the
Blandford--Znajek mechanism, magnetic flux threading a spinning BH
extracts its rotational energy as an electromagnetic outflow
\citep{blandford_znajek_1977}. By contrast, the Blandford--Payne mechanism
uses open magnetic-field lines anchored in the accretion disk to
centrifugally accelerate a matter-loaded outflow
\citep{blandford_payne_1982}. The accretion state nevertheless provides
a useful indication of when these mechanisms may operate.
Geometrically thick, radiatively inefficient flows occur at low
accretion rates, $\dot{M}\lesssim0.01\dot{M}_{\rm Edd}$, where cooling is
inefficient \citep{Muryel_lowedd_jet_21}, and at super-Eddington rates,
$\dot{M}\gtrsim\dot{M}_{\rm Edd}$, where photon trapping suppresses
radiative escape. In both regimes, the thick flow forms a polar funnel
that can accumulate large-scale magnetic flux and collimate outgoing
electromagnetic energy and matter. Although our hydrodynamical
simulations do not model magnetic jet launching directly, the component
accretion rates allow us to identify when neither, one, or both BHs
occupy an accretion state favorable for jet production.

To do so, we must first scale our numerical accretion rates, which are in units $M_{\rm{bin}}/\tau_b$, to Eddington units. For a BH of mass $M$, the Eddington accretion rate is given by
\begin{equation}\label{eqn:mdot_edd}
    \dot{M}_{\rm Edd} \equiv \frac{4\pi GM \mu_e m_p}{c \eta \sigma_t}
\end{equation}
where $m_p$ is the proton mass, $\mu_e$ is the mean molecular weight per electron ($\mu_e \simeq 0.6$ for ionized gas with solar abundances), $\sigma_t$ is a fiducial Thomson scattering cross-section, and $\eta \equiv L/(\dot{M}c^2) \simeq 0.1$ is the radiative efficiency. To retain information about the relative accretion rates of the two BHs, we set the binary accretion rate $\dot{M}_b \equiv \dot{M}_1 + \dot{M}_2$ equal to $\gamma \dot{M}_{\rm Edd}$ (with $\gamma$ an arbitrary scale factor evaluated at the total binary mass), and convert the accretion rate of each individual BH to its own Eddington units.

\begin{figure*}
    \centering
    \includegraphics[width=\textwidth]{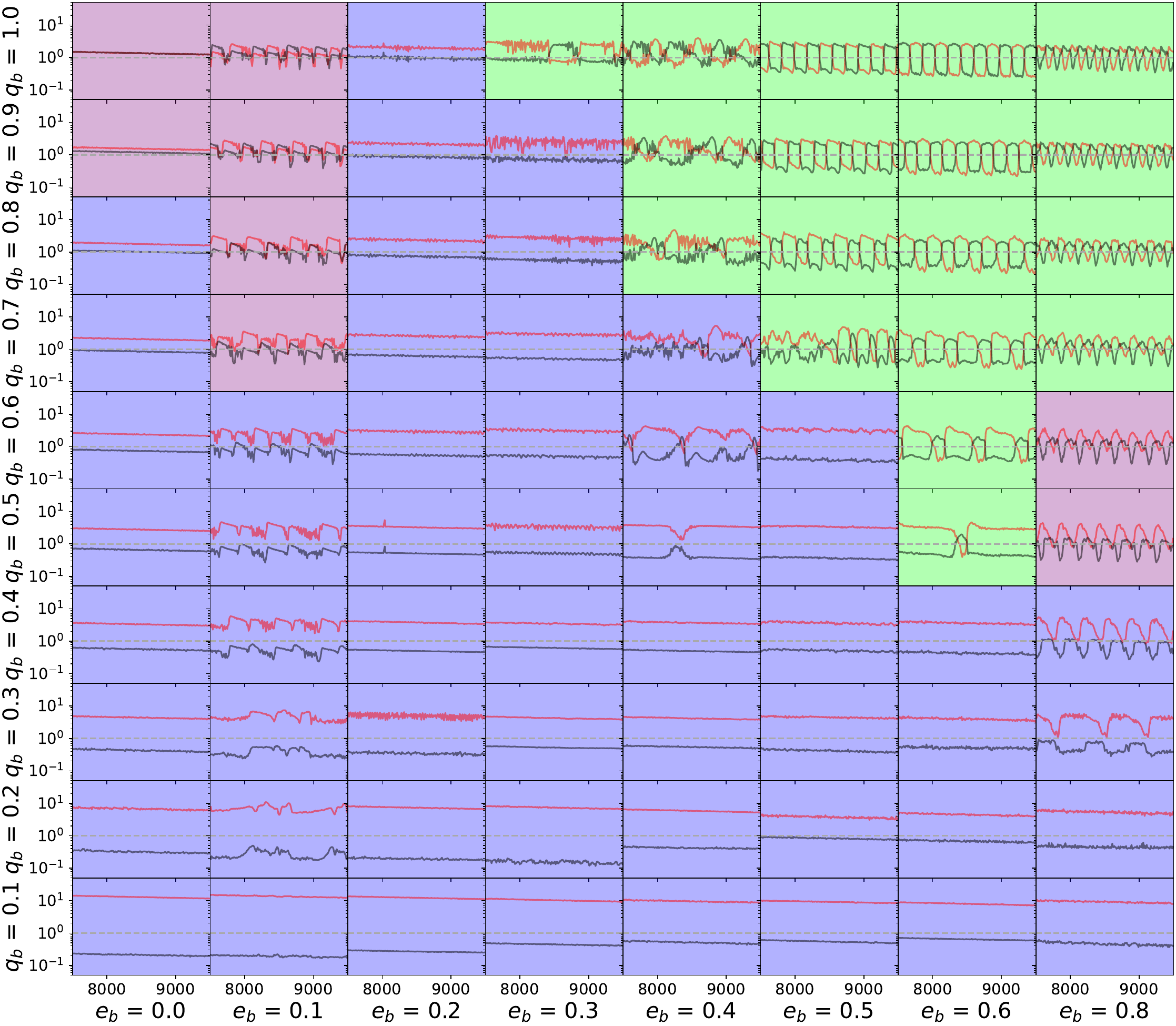}
    \caption{Illustration of Eddington-normalized accretion rates in our simulation suite, with panel backgrounds colored by jet-regime. The panel structure is the same as that of \autoref{fig:lambda_full}, with time on the x-axis of each panel and the individual per-BH accretion rates (in their respective Eddington units) on the y-axis. We display the accretion rate of the primary (black), the secondary (red), and the jet-launching threshold of $1\dot{M}_{\mathrm{Edd}}$ (horizontal dashed gray). Blue background colors indicates a single jet, purple indicates dual-jet regimes, and green indicates flickering jet regimes. At the fiducial $\dot{M}_b = 1.1\,\dot{M}_{\mathrm{Edd}}$ most binaries are single-jet (with the jet launched by the preferentially-accreting  and therefore super-Eddington secondary), while flickering jets are confined to high $e_b$, high $q_b$, and dual-jet cells are rare.}
    \label{fig:mdot_edd}
\end{figure*}

\begin{figure}
    \centering
    \includegraphics[width=1\linewidth]{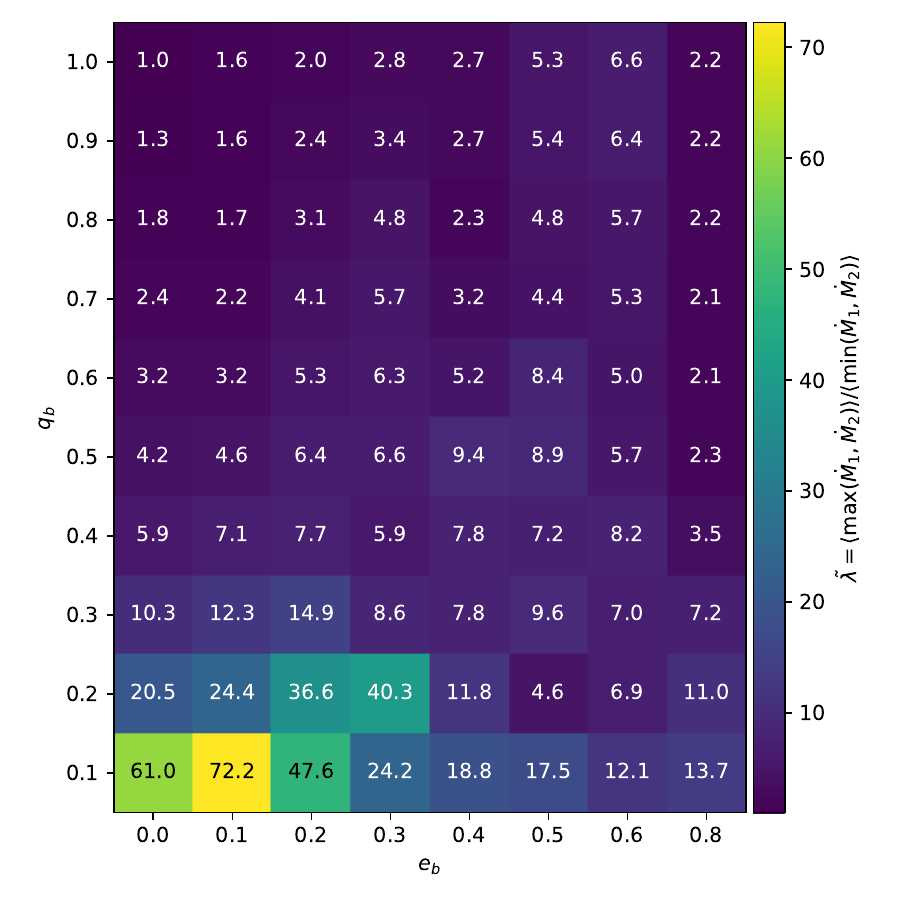}
    \caption{The accretion-rate ratio $\tilde{\lambda} \equiv \langle\max(\dot{M}_1, \dot{M}_2)\rangle / \langle\min(\dot{M}_1, \dot{M}_2)\rangle$ for our simulation suite, in $\dot{M}_{\rm Edd}$ units. The x-axis is the binary eccentricity $e_b$ and the y-axis is the binary mass-ratio $q_b$; cell colors and labels indicate $\tilde{\lambda}$ on a linear scale. Combined with an assumed binary accretion rate $\dot{M}_b$, $\tilde{\lambda}$ converts directly into the accretion rate of each BH, allowing the jet-launching regime (single, dual, or flickering; see \autoref{fig:mdot_edd}) to be predicted from $(e_b, q_b)$ alone for a given $\dot{M}_b$. $\tilde{\lambda}$ is largest at low $q_b$, where the per-BH rate disparity is most extreme, so the jet regime can be read off $(e_b, q_b)$ once $\dot{M}_b$ is fixed.}
    \label{fig:lambda_tilde_map}
\end{figure}

\subsubsection{Single, dual, and flickering jets}\label{sec:jet_zoo}

In \autoref{fig:mdot_edd} we set the accretion rate of the binary to be $1.1 \dot{M}_{\rm Edd}$ assuming a $10^{7} M_{\odot}$ binary and plot the accretion rate for both BHs normalized to their respective Eddington accretion rates. The horizontal gray dashed line is at $1\dot{M}_{\rm Edd}$ to represent the accretion rate above which jets are likely to launch. The red lines represent the accretion rate of the secondary, and the black lines are of the primary. The background color of the panel is associated with different jet-behaviors: purple for dual jets, blue for a single jet, green for flickering jets. The time-slice displayed is arbitrary and serves to merely highlight the accretion behavior.

A striking feature of \autoref{fig:mdot_edd} is the wide variety in magnitude between the two BHs' accretion rates. Since the Eddington rate scales linearly with the BH mass we expect that the accretion rates of the secondary to be increased greatly when normalized to Eddington units. This is evidenced by the nearly 2 order of magnitude difference between the secondary and primary at $q_b=0.1$. Further, we also notice that the behavior of the individual accretion rates of the black holes are quite varied, as the $\lambda(t)$ results suggested. Aside from the differences in whether the BHs' accretion rates are stable or not, the profile of the accretion rate itself is varied. Some binaries experience accretion rates that are close to sinusoidal (e.g $e_b=0.6, 0.8$), others seem to closer resemble square-waves (e.g.\ $e_b=0.3$ and $0.5$ at $q_b=1$), others yet have quite sharp breaks that evade simple characterizations (e.g.\ $(e_b, q_b) = (0.1, 0.7)$). Further, we note that the accretion rate of one BH is not always simply the accretion rate of the other with a different baseline and $\pi/2$ phase shift. Rather, they can take on notably different profiles from each other, resulting, at times, in both BHs experiencing a local peak in accretion rate, but because of different accretion rate amplitudes result in a peak in $\lambda$. It is this plethora of individual BH accretion rates, and the way in which they compare to each other, that yield an interesting assortment of jet-launching behaviors.

In \autoref{fig:mdot_edd} we delineate three broad regimes: a) binaries where one BH launches a jet (blue), b) binaries where both BHs coincidentally launch jets in a sustained and repeated fashion (purple), c) binaries where both BHs launch jets in a successive, alternating fashion (green). We describe these jet-behavior regimes as \emph{single jets}, \emph{dual jets}, and \emph{flickering jets}, respectively. We assigned jet-regimes by determining whether the accretion rate of each BH surpassed a threshold value of $1.1 \dot{M}_{\rm Edd}$ for greater than $50\,\tau_b$ (a threshold chosen modestly above unity to allow the disk thickness to inflate enough to support the funnel collimation discussed above) and whether those instances were temporally coincident for greater than $50\,\tau_b$. The $50\,\tau_b$ duration was chosen empirically: it is long enough to filter out short-lived threshold excursions (single-orbit transients, accretion bursts) and require a sustained launching episode, but short enough to preserve the alternating cadence we want to detect in the flickering regime. Jet activity is itself expected to follow the inner-disk dynamical time $t_{\rm dyn} \sim \Omega_K^{-1}$, which for our 2D setup is of order a binary orbital period; the $50\,\tau_b$ window therefore samples many dynamical times. These thresholds (the $1.1\,\dot{M}_{\rm Edd}$ amplitude, the $50\,\tau_b$ duration, and the $0.05$ normalized-FFT-amplitude cut used for the period extraction in \S\ref{sec:preliminary_analysis}) are heuristic, and cells sitting near a threshold should be read as marginal; modest changes to these values (e.g.\ a normalized-amplitude cut of $0.03$ or $0.07$) would reassign borderline cells, but the broad single/dual/flickering partition is set by the large per-BH accretion-rate disparities rather than by the precise cut.

At the fiducial binary accretion rate $\dot{M}_b = 1.1\,\dot{M}_{\rm Edd}$, the two BHs must divide a single, fixed Eddington budget. Because each BH's own Eddington rate scales with its mass, the two per-BH Eddington rates sum to that of the binary ($\dot{M}_{\rm Edd,1} + \dot{M}_{\rm Edd,2} = \dot{M}_{\rm Edd,b}$): the more one BH exceeds its own Eddington limit, the less of the shared budget is left for the other. The two therefore cannot both sit \textit{well} above the $1.1\,\dot{M}_{\rm Edd}$ jet threshold at the same time. Most cells are accordingly single-jet systems, with the jet launched by the preferentially-accreting secondary, whose Eddington-normalized rate is boosted relative to the primary, most strongly at low $q_b$. The marginal exception is the near-equal-mass, low-$e_b$ corner: there the two BHs split the budget almost evenly, so each sits at $\dot{M}_i \approx 1.1\,\dot{M}_{{\rm Edd},i}$, right at the jet threshold, where dual jets are at best marginal. No cell is left jetless, however: at this near-Eddington rate at least one BH clears the threshold in every $(e_b, q_b)$ cell of \autoref{fig:mdot_edd}, so a binary accreting near its Eddington limit always launches \textit{at least} one jet.

The flickering jet systems are clustered at higher $q_b$ and $e_b$. Due to large $\lambda$ amplitudes, such systems are able to exist up to $q_b=0.5$. It is important to note that since these flickering jets are dependent on large $\lambda$ oscillations they are unique to $e_b \neq 0$ binaries.

A handful of cells are classified as dual-jet at $\dot{M}_b = 1.1\,\dot{M}_{\rm Edd}$, chiefly the near-equal-mass $e_b = 0$ binaries, where both BHs accrete at a nearly equal, stable rate close to Eddington. As noted above, these are marginal cases sitting right at the threshold. Robust, sustained dual jets (with both BHs comfortably above threshold) are instead expected when the binary total lies either well above Eddington (both super-Eddington) or deep in the ADAF regime (both radiatively inefficient at $\dot{M} < 0.01\,\dot{M}_{\rm Edd}$, which can also drive jets), as we illustrate for two bracketing values of $\dot{M}_b$ in \autoref{fig:mini_jet_regimes}.

\subsubsection{Prevalence of Flickering Jets}\label{sec:prev_flicker}
While dual jets from BBH systems have been suggested before \citep{Palenzuela_dualjet_10,BaumgarteShapiro2011,Qian_dualjet_19} and demonstrated numerically in a range of GRMHD setups \citep{Gold_2014_jets, Gutierrez_24, Ressler_dualjet_25, Ruiz_Shapiro_23, Most_Wang_24, Ennoggi_25}, we believe that we are the first to identify flickering jets as a distinct observational regime and systematically predict where they should occur across binary parameter space. Although our hydrodynamical simulations do not model magnetic jet launching directly, recent GRMHD simulations provide strong support for this interpretation. In particular, \citet{combi2026dualjets} find that asymmetric feeding from an eccentric circumbinary disk causes the magnetic flux and jet luminosity to alternate between the two BHs, producing an on--off dual-jet state. Their calculation directly demonstrates this behavior for a circular, equal-mass binary, while our results predict its prevalence across binary mass ratio, eccentricity, and total accretion rate.

\autoref{fig:lambda_tilde_map} displays $\tilde{\lambda}$,
\begin{equation}
    \tilde{\lambda} \equiv \frac{\langle \max (\dot{M}_{1}, \dot{M}_{2}) \rangle }{\langle \min (\dot{M}_{1}, \dot{M}_{2}) \rangle},
\end{equation}
the ratio of the maximum and minimum accretion rates of the binary components. Combined with an assumed binary accretion rate $\dot{M}_b$, $\tilde{\lambda}$ fixes the per-BH rate via
\begin{equation}
    \dot{M}_1 = \frac{\dot{M}_b}{\tilde{\lambda} + 1},
\end{equation}
and so predicts the jet-launching regime (single, dual, or flickering) from $(e_b, q_b)$ alone for any $\dot{M}_b$. Given the clustering of jet behavior in $(e_b, q_b)$ parameter space, determination of whether a system sustains dual jets, flickering jets, or a single jet could greatly constrain the orbital parameters of the system.

\begin{figure*}
    \centering
    \includegraphics[width=1\linewidth]{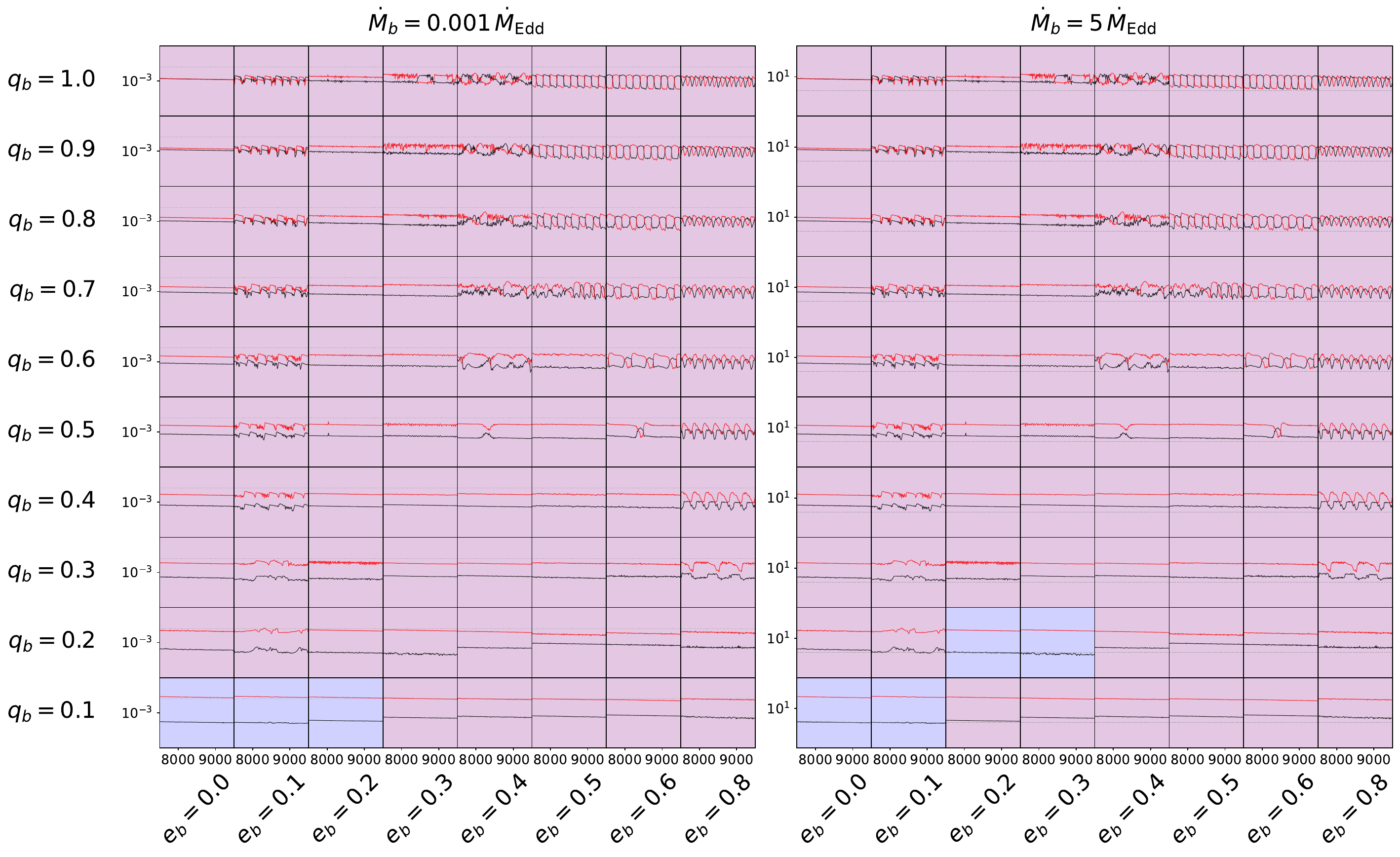}
    \caption{Reproduction of \autoref{fig:mdot_edd} at two bracketing values of the binary accretion rate that place each BH inside one of the two jet-launching regimes discussed in \S\ref{sec:observations}: a deep-ADAF case $\dot{M}_b = 0.001\,\dot{M}_{\mathrm{Edd}}$ (left, applying the $\dot{M} < 0.01\,\dot{M}_{\mathrm{Edd}}$ ADAF threshold) and a mildly super-Eddington case $\dot{M}_b = 5\,\dot{M}_{\mathrm{Edd}}$ (right, applying the $\dot{M} > 1.1\,\dot{M}_{\mathrm{Edd}}$ thick-disk threshold). Each cell shows the Eddington-normalized accretion rates of the primary (black) and secondary (red) on a log y-axis (range shifted per panel to match the data); the dashed gray horizontal marks the $\dot{M}_i = \dot{M}_{\mathrm{Edd},i}$ super-Eddington threshold and the dotted gray horizontal marks the $\dot{M}_i = 0.01\,\dot{M}_{\mathrm{Edd},i}$ ADAF threshold. Backgrounds are colored by jet regime: blue (single-jet, one BH meets its panel's threshold) and purple (dual-jet, both BHs meet their threshold simultaneously sustained). At both extremes dual jets dominate; only single-jet (blue) and dual-jet (purple) cells are realized. The fiducial $\dot{M}_b = 1.1\,\dot{M}_{\mathrm{Edd}}$ shown in \autoref{fig:mdot_edd} represents the radiatively-efficient transition band where the regime mix is richest. Away from the fiducial rate dual jets dominate and flickering disappears; the surviving single jets are launched by the primary in the deep-ADAF (left) panel and by the secondary in the super-Eddington (right) panel.}
    \label{fig:mini_jet_regimes}
\end{figure*}

To illustrate the dependence on $\dot{M}_b$, \autoref{fig:mini_jet_regimes} reproduces \autoref{fig:mdot_edd} at two bracketing values: a deep-ADAF case $\dot{M}_b = 0.001\,\dot{M}_{\rm Edd}$ (left) and a mildly super-Eddington case $\dot{M}_b = 5\,\dot{M}_{\rm Edd}$ (right). At both extremes, dual jets dominate the parameter space, with single-jet cells surviving only at low $q_b$, precisely the region with the largest $\tilde{\lambda}$ in \autoref{fig:lambda_tilde_map}, where the rate discrepancy is severe enough that only one BH can cross any jet threshold regardless of $\dot{M}_b$. Which BH that is, however, flips between the two extremes: in the deep-ADAF panel the single jet is launched by the \textit{primary} (the starved component that alone falls below the radiatively-inefficient $0.01\,\dot{M}_{\rm Edd}$ threshold), whereas in the super-Eddington panel, as at the fiducial rate, it is the preferentially-accreting \textit{secondary} that alone exceeds the super-Eddington threshold. This distinction is observationally relevant: a single jet anchored to the more massive primary sits closer to the binary's center of mass and sweeps out a smaller orbit, so its launching point wobbles less than that of a secondary-anchored jet. The fiducial $\dot{M}_b = 1.1\,\dot{M}_{\rm Edd}$ in \autoref{fig:mdot_edd} sits in the transition band where the regime mix is richest; pushing $\dot{M}_b$ well above or below it drives most of the parameter space into the dual-jet regime and removes the flickering cells. Robust dual jets are thus confined to these two extremes: across the intermediate range $\dot{M}_b \sim 0.01$--$1\,\dot{M}_{\rm Edd}$, which brackets the Eddington ratios of order $0.1$ typical of luminous quasars, the shared Eddington budget keeps the second BH below threshold in essentially every cell. A bright-quasar SMBBH accreting near these rates should therefore display a single jet (or, at high $e_b$ and $q_b$, flickering) rather than two simultaneous jets; sustained dual jets instead point to a binary that is either strongly super-Eddington or in the deep-ADAF regime.

\subsubsection{Observing a flickering jet}

\begin{figure*}
    \centering
    \includegraphics[width=\textwidth]{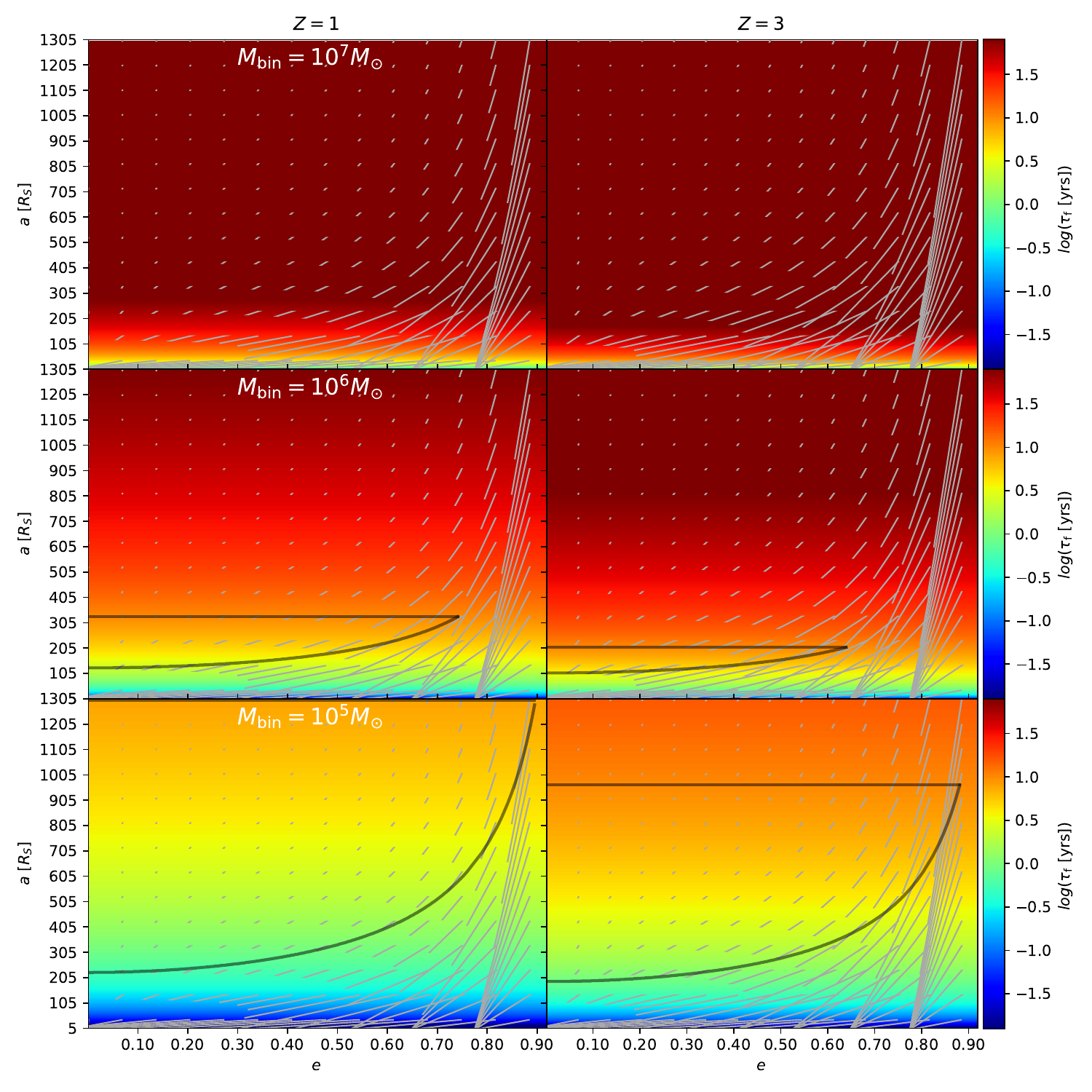}
    \caption{Region of $(e_b, a_b)$ parameter space in which a system satisfies the flickering-jet observability constraints (a flicker time $\tau_f \le 10$~yr in the observer frame, and a remaining merger time $\ge 100$~yr) for various binary masses and redshifts. The black contours enclose the detection-friendly region in each panel. The x-axis of each panel is the binary eccentricity $e_b$, the y-axis is the binary semi-major axis $a_b$ in Schwarzschild radii. The panel fill colour shows $\log_{10}(\tau_f / {\rm yr})$, the observer-frame flicker time: blue marks a short (sub-year) flicker time and red a long one ($\gtrsim 10$~yr), so the detection-friendly region (black contour) hugs the blue/green low-$\tau_f$ part of each panel. The gray line segments depict the net change in $(e_b, a_b)$ a binary undergoes over 10 orbits of GW emission. The flicker-observability window covers substantial regions of $(e_b, a_b)$ for $10^5$--$10^6\,M_\odot$ binaries out to $z = 3$, but is essentially closed for $10^7\,M_\odot$ binaries, whose longer flicker time exceeds $10$~yr.}
    \label{fig:timescales}
\end{figure*}

"Flickering jets" are a potentially distinctive electromagnetic signature of BBHs: unlike the stochastic variability of a single AGN jet, flickering produces an alternating, quasi-periodic switching of jet activity between two spatially offset launching sites (the two BHs), which, if the jet orientations differ, could in principle be distinguished from ordinary single-source AGN variability. A suggestive observational analogue is the SMBHB candidate PKS 2131-021, whose jet-associated sinusoidal radio modulation disappeared for approximately two decades before returning with the same period and phase \citep{O_Neill_2022}. We caution, however, that because the preferentially-accreting BH lies above threshold in most cells, a flickering binary will typically have at least one jet active at any given time; if the two jets are closely aligned, this alternation can resemble the ordinary stochastic flickering of a single AGN jet, and the binary nature is cleanest to recover when the two jet orientations (set by the individual BH spins) differ appreciably. In order to use them to find BBH systems, we must ensure they "flicker" (i.e. switch which BH is preferentially accreting) on a humanly trackable time-scale. In the following we compute  and place constraints on the time to "flicker".

Firstly, we require that the time to flicker $\tau_f$ be fewer than $10$ years in the observer's rest-frame, so that a few cycles could be possible to find on a humanly trackable time-scale. For simplicity we adopt a representative flickering period of $300\,\tau_b$ (at the binary's initial orbital period), of order the few-hundred-$\tau_b$ CBD apsidal precession period that paces the $\lambda(t)$ oscillations (\S\ref{sec:preliminary_analysis}). We also require that the binary not merge in less than $100$ years, in order to ensure that these systems are not exceedingly rare.

In \autoref{fig:timescales} we display the region of parameter space that satisfies the above time constraints (within black lines) for various binary masses at various redshifts. The x-axis of each panel is the eccentricity $e_b$, the y-axis is the binary semi-major axis $a_b$ in Schwarzchild radii, and $\tau_f$ is reported in the observer frame. The gray lines represent the change in eccentricity and semi-major axis for the binary due to 10 orbits worth of GW radiation, computed via \citet{peters_64}.

\autoref{fig:timescales} shows that the flicker-observability window depends strongly on binary mass. The $10^5\,M_\odot$ and $10^6\,M_\odot$ binaries both retain considerable regions of $(e_b, a_b)$ parameter space satisfying the constraints out to $z = 3$, with the window shrinking as redshift increases (more rapidly for the $10^6\,M_\odot$ case). Because the flicker time $\tau_f = 300\,\tau_b$ grows with binary mass, lighter binaries flicker fastest and retain an observability window to higher redshift; the $10^5\,M_\odot$ window remains large across both redshifts shown. By contrast, the $10^7\,M_\odot$ binaries flicker too slowly to satisfy $\tau_f \le 10$~yr over essentially the entire parameter space; their window has already closed by $z \approx 1$ and is absent from both panels. The persistence of a sizeable observability window for $10^5$--$10^6\,M_\odot$ binaries out to $z = 3$ provides encouraging evidence that flickering-jet systems could be detected.

In addition to observing a flicker occur, we note that jets are extended emission sources and thereby provide us an ability to deduce evidence of a past flicker. If we could determine a geometric separation in the structure of a helical jet, this could indicate that the emission is from a binary that flickered in the past. 

Beyond jets, time-variable preferential accretion also affects binary light-curve signatures. Self-lensing flares depend on the relative brightnesses of the two minidiscs, which determine which black hole acts as the lens and which as the source. Using matched filters with \textit{binlite} templates, \citet{Park2026SLF5} recover the binary period and inclination but obtain weaker constraints on eccentricity, principally because an inadequately sampled secondary flare produces degeneracies among ($e_{\rm b}$), the argument of periapsis, and the overall orbital phase. They also identify a separate hydrodynamic-template limitation: \textit{binlite} assigns a single time-averaged accretion-rate ratio, even though its templates are extracted from a simulation in which eccentricity is swept continuously. Adjacent eccentricities therefore correspond to different simulation times and can reverse which minidisc is brighter. Our finding that $\lambda(t)$ is modulated on the circumbinary disk’s apsidal-precession timescale provides a physical interpretation of this behavior: part of the rapid variation attributed to $e_{\rm b}$ may instead arise because nei        ghbouring templates sample different disk-precession phases. Future template banks should therefore condition the accretion ratio on precession phase rather than treating it as a deterministic function of eccentricity. This would prevent disk-phase variability from being misidentified as intrinsic eccentricity dependence. The same issue affects Doppler-boost models \citep{dorazio_spikey, charisi_doppler} that assume a fixed accretion-rate ratio throughout the light curve.

\subsection{Unequal-mass sources}\label{sec:unequal_mass}

In addition to affecting jet production, gas accretion determines the mass ratio that a binary carries into the LISA band. An interesting question is whether a binary that is initially unequal can retain some inequality during its gas-driven evolution. As shown in \autoref{fig:qdot_heatmap}, binaries with $q_b<1$ generally evolve toward equal mass, but the evolution becomes slow at high $q_b$ and high $e_b$. We therefore evolve a representative sample of binaries with initial mass ratio $q_{b,0}=0.8$ and $0.9$ to determine its mass ratio evolution.

Gas accretion changes the binary semi-major axis, eccentricity, mass ratio, and total mass. We take the gas-driven $\dot a_{\rm gas}$ and $\dot e_{\rm gas}$ measured by \Stwentythree, together with the time-averaged $\langle\dot q_b\rangle$ measured in \autoref{fig:qdot_heatmap}. These rates are linearly interpolated across the simulated $(e_b,q_b)$ grid and scaled to the adopted binary accretion rate,
\begin{equation}
\dot m \equiv \frac{\dot M_b}{\dot M_{\mathrm{Edd}}}.
\end{equation}

Gravitational-wave emission also shrinks and circularizes the binary. We include these effects using the standard orbit-averaged quadrupole expressions \citep{peters_64},
\begin{align}
\dot{a}_{\rm GW} &=
-\frac{64G^3M_b^3q_b}
{5c^5a_b^3(1+q_b)^2}
f(e_b),\\
\dot{e}_{\rm GW} &=
-e_b\frac{304G^3M_b^3q_b
\left(1+\frac{121}{304}e_b^2\right)}
{15c^5a_b^4(1-e_b^2)^{5/2}(1+q_b)^2},
\end{align}
where
\begin{equation}
f(e_b)=
\frac{1+\frac{73}{24}e_b^2+\frac{37}{96}e_b^4}
{(1-e_b^2)^{7/2}}.
\end{equation}

The coupled evolution is therefore
\begin{equation}
\begin{aligned}
\dot a_b &=
\dot a_{\rm gas}(e_b,q_b)
+\dot a_{\rm GW}(a_b,e_b,q_b,M_b),\\
\dot e_b &=
\dot e_{\rm gas}(e_b,q_b)
+\dot e_{\rm GW}(a_b,e_b,q_b,M_b),\\
\dot q_b &=
\langle\dot q_b\rangle_{\rm gas}(e_b,q_b).
\end{aligned}
\label{eqn:num_integ}
\end{equation}
We numerically integrate \autoref{eqn:num_integ} from the chosen initial values $a_{b,0}$, $e_{b,0}$, and $q_{b,0}$. The binary mass grows at the adopted rate $\dot M_b=\dot m \,\dot M_{\mathrm{Edd}}$ and is updated throughout the integration. Gravitational-wave emission changes $a_b$ and $e_b$ but leaves $q_b$ unchanged at this order.

Before viscous decoupling, the gas and gravitational-wave terms are integrated simultaneously. After decoupling, we switch off the gas contributions to $\dot a_b$ and $\dot e_b$ but allow differential accretion to continue while gas remains available. Simulations indicate that accretion is not abruptly suppressed at decoupling \citep{farris_2015,krauth2023b,Ennoggi2025}. In practice, whether differential accretion is stopped at decoupling or allowed to persist has a negligible effect on $q_{b,\mathrm{LISA}}$ for the systems considered here.

Following the characteristic-strain construction used in \citet{DeLaurentiis24}, we define LISA entry as the first upward crossing of the $n=2$ characteristic strain and the sky-averaged LISA characteristic-noise curve. The observed frequency is $f=2f_{\mathrm{bin}}/(1+z)$, and
\begin{equation}
h_{c,2}(f)=\frac{1}{\pi D_L}
\left[
\frac{2(G\mathcal{M})^{5/3}(2\pi)^{2/3}g_2(e_b)}
{3F(e_b)[(1+z)f]^{1/3}c^3}
\right]^{1/2},
\end{equation}
where $\mathcal{M}=M_bq_b^{3/5}(1+q_b)^{-6/5}$. We identify the crossing through $h_{c,2}(f)=\sqrt{fS_n(f)}$, using the \citet{robson2019} LISA sensitivity curve.

\begin{figure*}
\centering
\includegraphics[width=\textwidth]{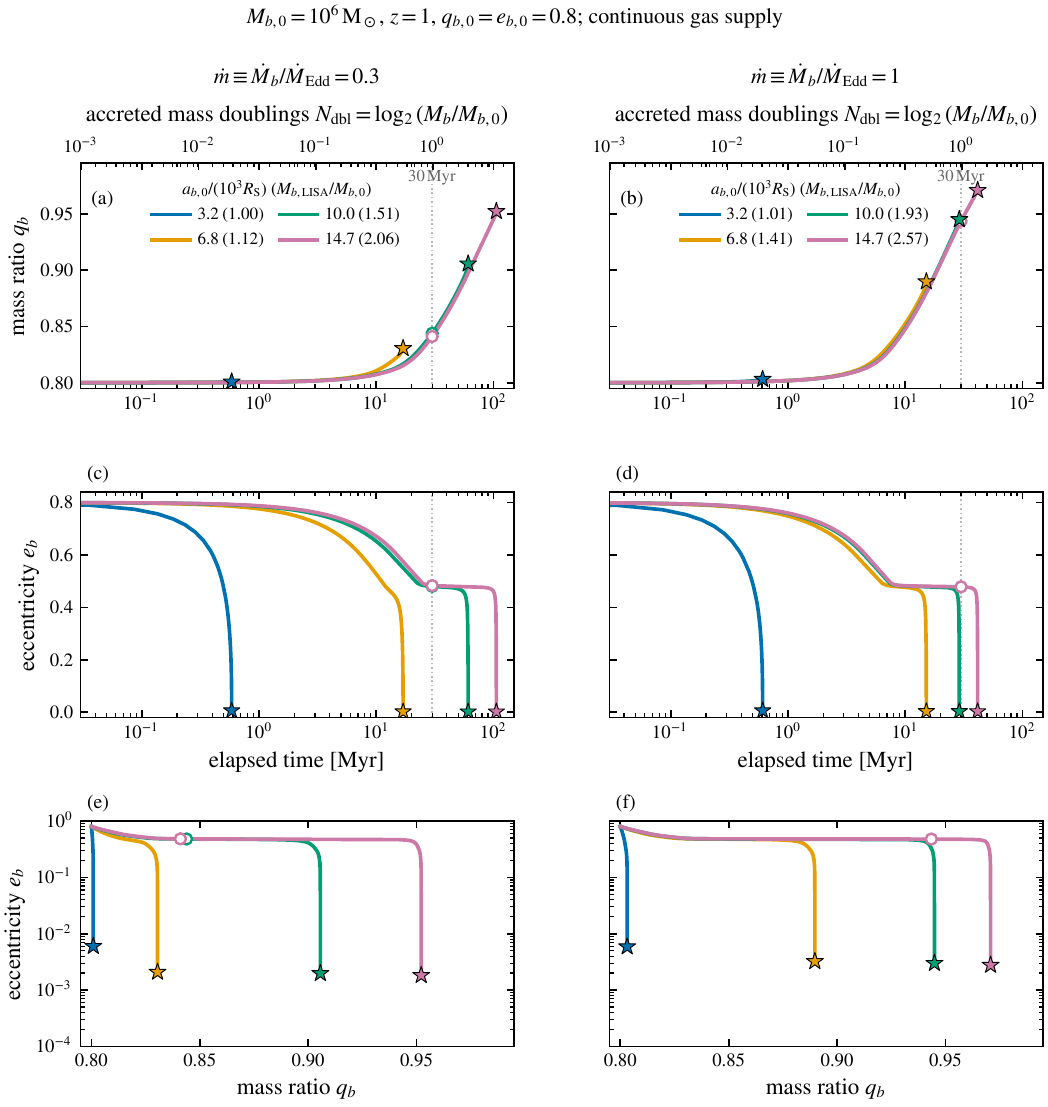}
\caption{Illustrative evolution of binaries with initial mass $M_{b,0}=10^6\,\mathrm{M}_\odot$, mass ratio $q_{b,0}=0.8$, and eccentricity $e_{b,0}=0.8$ at $z=1$, for $\dot M_b/\dot M_{\mathrm{Edd}}=0.3$ (left) and $1$ (right). Colors denote the initial separation $a_{b,0}$. The upper-panel legends give $a_{b,0}/(10^3R_{\mathrm{S}})$, with $M_{b,\mathrm{LISA}}/M_{b,0}$ in parentheses. The upper and middle rows show $q_b(t)$ and $e_b(t)$, while the lower row shows the trajectory through $(q_b,e_b)$ space on a logarithmic eccentricity axis. Gas and gravitational-wave terms are integrated simultaneously. Open circles mark a source-frame time of $30\,\mathrm{Myr}$ when reached before LISA entry. Stars mark the first upward crossing of the $n=2$ characteristic strain and the \citet{robson2019} characteristic-noise curve. The $30\,\mathrm{Myr}$ marker is diagnostic and is not an imposed gas-supply cutoff. The results suggest that \textit{LISA} can constrain the accretion history of binaries.}
\label{fig:lisa_qe_trajectories}
\end{figure*}

\autoref{fig:lisa_qe_trajectories} shows representative $M_{b,0}=10^6\,\mathrm{M}_\odot$ binaries at $z=1$ with $q_{b,0}=e_{b,0}=0.8$. Systems beginning closer to the \textit{LISA} band accrete very little and enter with essentially their initial mass ratio. Systems beginning farther out remain coupled to the gas for longer, gain more mass, and evolve toward more equal mass. For $\dot m=0.3$, the representative tracks enter with $q_{b,\mathrm{LISA}}=0.801$ and $0.952$ after growing by factors of $1.004$ and $2.059$. For $\dot m=1$, they enter with $q_{b,\mathrm{LISA}}=0.803$ and $0.971$ after growing by factors of $1.014$ and $2.571$. Their eccentricities at the sensitivity-curve crossing span approximately $1.8\times10^{-3}$ and $6.0\times10^{-3}$.

The open circles in \autoref{fig:lisa_qe_trajectories} mark a source-frame time of $30\,\mathrm{Myr}$ when that time is reached before LISA entry. This is a diagnostic marker rather than an assumed disk lifetime. The stars show the limiting continuous-supply case in which gas remains available until the sensitivity-curve crossing. Because continuous supply maximizes the available mass growth, these endpoints provide an upper limit on how strongly accretion can drive the binaries toward equal mass.

Across the broader sweep over $e_{b,0}=0.4$ and $0.8$ and the four sampled initial separations, binaries with $q_{b,0}=0.8$ enter with $q_{b,\mathrm{LISA}}=0.801$ and $0.963$ for $\dot m=0.3$ and $0.803$ and $0.977$ for $\dot m=1$. Binaries with $q_{b,0}=0.9$ enter with $q_{b,\mathrm{LISA}}=0.901$ and $0.984$ and $0.902$ and $0.990$, respectively. Thus, evolution toward equal mass is common, but a pile-up exactly at $q_b=1$ is not inevitable.

This result is consistent with \citet{valli2024}, who showed that substantial orbital and mass-ratio evolution generally requires a gas reservoir comparable to the binary mass. In our calculation, the tracks that accrete the most mass also evolve closest to unity, whereas binaries that encounter the LISA sensitivity curve after little mass growth retain nearly their initial $q_b$. It is also consistent with the asymptotic evolution toward unity found for circular binaries by \citet{xu2026}. The additional result here is to couple the measured eccentric-binary mass-ratio evolution to the simultaneous evolution of $a_b$ and $e_b$ and follow it to a sensitivity-based LISA endpoint. The calculation does not predict the LISA population because the distributions of initial separations, eccentricities, accretion rates, and gas-active lifetimes remain uncertain. It instead demonstrates that appreciably unequal binaries can survive into the LISA band for physically reasonable accretion histories.

\section{Summary and conclusions}\label{sec:conclusion}
This paper has provided the most extensive report to date on preferential accretion $\lambda \equiv \frac{\dot{M}_2}{\dot{M}_1}$ and mass ratio rate of change $\dot{q}_b$ for SMBBHs embedded in thin prograde CBDs. We provide insight into the behavior of these quantities over time and their dependence on $q_b$ and $e_b$. We also conduct a preliminary investigation into how the CBD regulates preferential accretion. We summarize our key findings below.

\begin{enumerate}
    \item Across $q_b$ and $e_b$, $\lambda(t)$ can be split into constant and time-varying regimes (\autoref{tab:stable_varying_grid}), broadly mirroring the split of the CBD into locked and precessing states.
    \item The time-averaged value $\langle \lambda \rangle$ is largest for low-$e_b$, low-$q_b$ binaries and declines toward higher $e_b$ and $q_b$, following the $q_b^{-0.9}$ trend of \Stwentythree at low eccentricity (\autoref{fig:lambda_mean_heatmap}).
    \item $\sigma_{\lambda}$ tracks the cavity eccentricity: both increase with $e_b$ and peak near $e_b = 0.6$ (\autoref{fig:lambda_std_heatmap}, compare \autoref{fig:a_e_cav_heatmap}).
    \item Across precessing systems in our suite, the CBD apsidal precession period and the $\lambda(t)$ oscillation period are equal. A direct cross-correlation of $\lambda(t)$ with the cavity-wall distance is strong in these precessing $(e_b,q_b)$ parameter combinations (median peak $\approx 0.9$) but occurs at a small non-zero lag rather than at the naively expected $\pi$ phase offset (Appendix~\ref{sec:appendix_crosscorr}). Cavity-wall distance therefore paces most of the variability in the precessing regime, while additional physics operates for the locked or irregular $(e_b,q_b)$ parameter combinations.
    \item We do not find evidence that a BH must be closer to the CBD cavity than its companion to accrete at a higher rate.
    \item Normalized to Eddington accretion rates, $\lambda(t)$ results in disparate accretion regimes for each BH in the binary, leading to unique jet-launching regimes. We delineate binaries that are likely to launch a sustained jet from one BH (single-jet), from each BH (dual-jet), or alternate in which BH launches a jet (flickering-jet).
    \item Mass-ratio evolution under gas accretion varies greatly across $q_b$ and $e_b$, and is particularly slow for high-$e_b$ and high-$q_b$ binaries. Namely, by coupling our measured $\langle \dot{q}_b \rangle$ to gas-driven orbital evolution and GW emission, we find that binaries beginning at $q_{b,0}=0.8$ retain $q_{b,\mathrm{LISA}}\simeq0.8$–$0.95$ after a fiducial $30\,\mathrm{Myr}$ episode at $\dot{M}=0.3\dot{M}_{\mathrm{Edd}}$–$1$.

\end{enumerate}

While our work has shed light on one aspect of the SMBBH--CBD system,
it is based on simplified physics in two-dimensional hydrodynamical
simulations. Future work should extend this study to three-dimensional
magnetohydrodynamical and radiative simulations, incorporate black-hole
spin, and allow the binary orbit and component masses to evolve
self-consistently. Retrograde CBDs also warrant separate investigation:
they produce qualitatively different orbital evolution, including
systematic binary hardening and eccentricity excitation, and develop
distinct inner-disk structures during the GW-driven inspiral
\citep{oneill_2025}. 
Such calculations will be
particularly important near $q_b=1$, where the identities of the
primary and secondary can interchange, and for determining how long
gas-driven evolution persists during the transition to the
gravitational-wave-dominated regime.

We conclude by noting that the mechanism behind preferential accretion is more complex than the near-uniform picture in which the secondary steadily out-accretes the primary along a smooth $q_b^{-0.9}$ trend \citep{farris_2014, duffell_dorazio_2020, siwek_prefacc}. Our results instead suggest that the precessing or locked, lopsided CBD regulates both the instantaneous partition of accretion and the long-term evolution of the binary mass ratio. Although binaries with $q_b<1$ generally evolve toward equal mass, this evolution becomes slow at large $q_b$ and large $e_b$. Our coupled gas and gravitational-wave integrations show that representative binaries beginning at $q_{b,0}=0.8$ can enter the LISA band with $q_{b,\mathrm{LISA}}\simeq0.80$--$0.95$ for $\dot m=0.3$ and $q_{b,\mathrm{LISA}}\simeq0.80$--$0.97$ for $\dot m=1$. Thus, the SMBBH population need not pile up exactly at $q_b=1$, and LISA mass-ratio measurements may retain information about the preceding CBD-driven phase. This suggests that \textit{LISA} may be able to place unique constraints on the gas accretion histories of SMBBHs. A complementary electromagnetic signature of the same cavity-regulated accretion may be the flickering jet regime identified above.

\section*{Acknowledgements}
The authors thank Roman Rafikov for his thoughts and feedback on the work. The authors thank the anonymous referees for helpful comments. ZH acknowledges support from NASA ATP grant 80NSSC22K0822 and LISA Preparatory Science grant 80NSSC24K0440. 
MS acknowledges support from Simons Foundation International grant SFI-MPS-SFJ-00006123. 
We acknowledge computing resources from Columbia University's Shared Research Computing Facility, in particular the Ginsburg HPC cluster.

\section*{Data Availability}
The data underlying this article will be shared on reasonable request to the corresponding author.

\bibliographystyle{mnras}
\bibliography{main}

\appendix

\section{Cross-correlation of the accretion-rate ratio and cavity-wall distance}\label{sec:appendix_crosscorr}

To test how tightly the time-variability of preferential accretion tracks the cavity geometry, we measure the normalized, lagged cross-correlation $C_{\lambda, r_2}(\Delta t)$ (\autoref{eqn:crosscorr}) between the accretion-rate ratio $\lambda(t)$ and the secondary--cavity-wall distance $r_2(t)$. Both time-series have the initial $3000\,\tau_b$ transient removed and are standardized to zero mean and unit variance before the correlation is computed. We restrict the analysis to $r_2$: because $r_1$ and $r_2$ are tightly anti-correlated proxies for the same cavity orientation (\S\ref{sec:preliminary_analysis}), $r_1$ carries no independent information. For each simulation we extract the principal peak $\max_{\Delta t} C_{\lambda, r_2}$ (the maximum of $C_{\lambda, r_2}(\Delta t)$ within $\pm$ half a precession period of zero lag) and the lag $\Delta t$ at which it occurs. Because the lag is only meaningful where the correlation is strong, we map the peak correlation across the full suite (\autoref{fig:cc_heatmap}) and summarize the lag by its median over the strongly-correlated cells (reported in \S\ref{sec:preliminary_analysis}), rather than mapping every cell's lag.

\begin{figure*}
    \centering
    \includegraphics[width=1\textwidth]{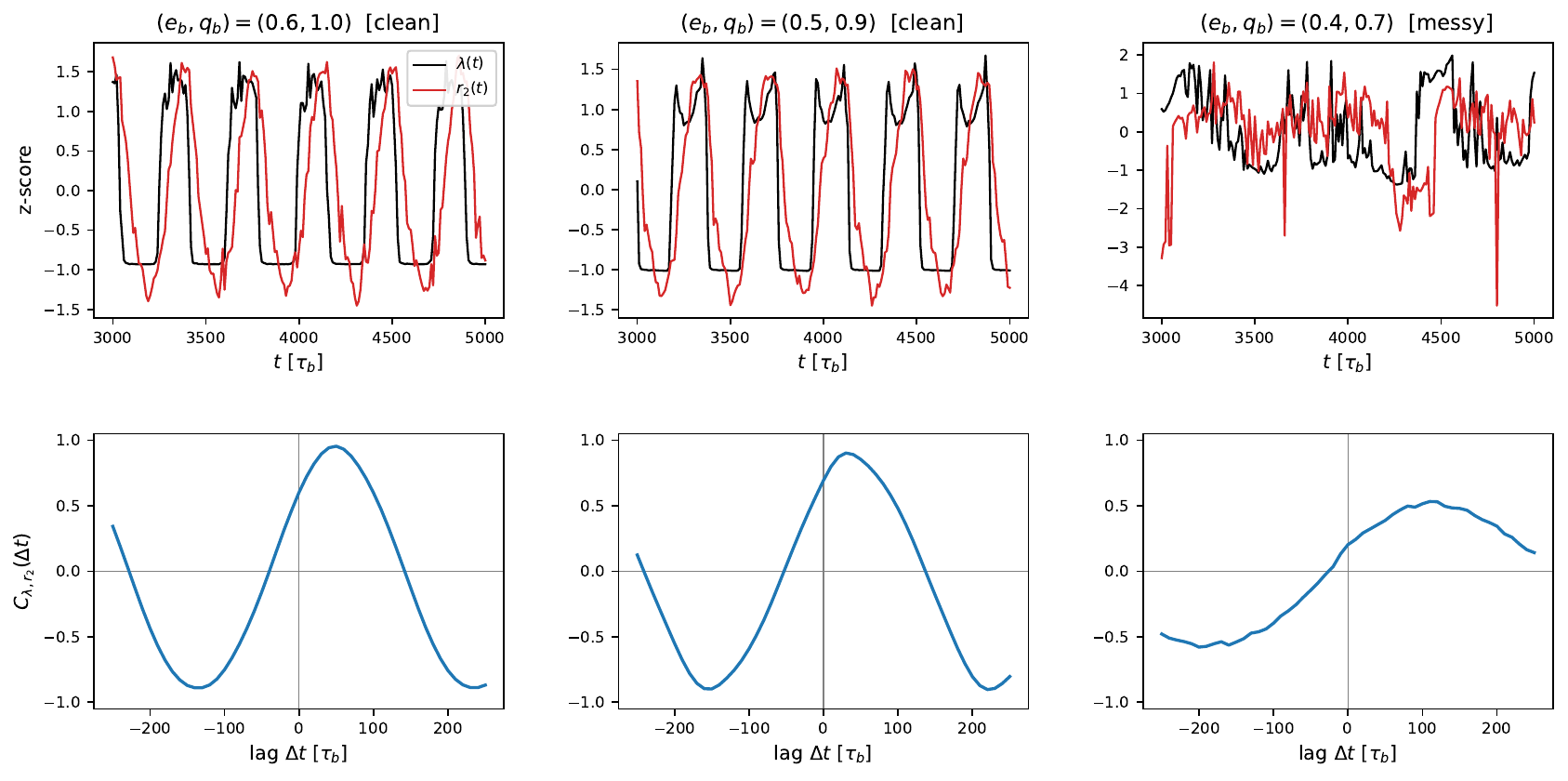}
    \caption{Lagged cross-correlation between the accretion-rate ratio $\lambda(t)$ and the secondary--cavity-wall distance $r_2(t)$ for three representative simulations. \textit{Top row:} the standardized (z-scored) $\lambda(t)$ (black) and $r_2(t)$ (red) over the window $3000 \le t/\tau_b \le 5000$. \textit{Bottom row:} the normalized cross-correlation $C_{\lambda, r_2}(\Delta t)$ (\autoref{eqn:crosscorr}). The two precessing cells (left, center) show a strong correlation peaking at a small positive lag; the messy cell (right) shows a weak, broad correlation. In the precessing regime $\lambda$ tracks the cavity-wall distance closely but with a finite response lag.}
    \label{fig:cc_examples}
\end{figure*}

\begin{figure}
    \centering
    \includegraphics[width=1\linewidth]{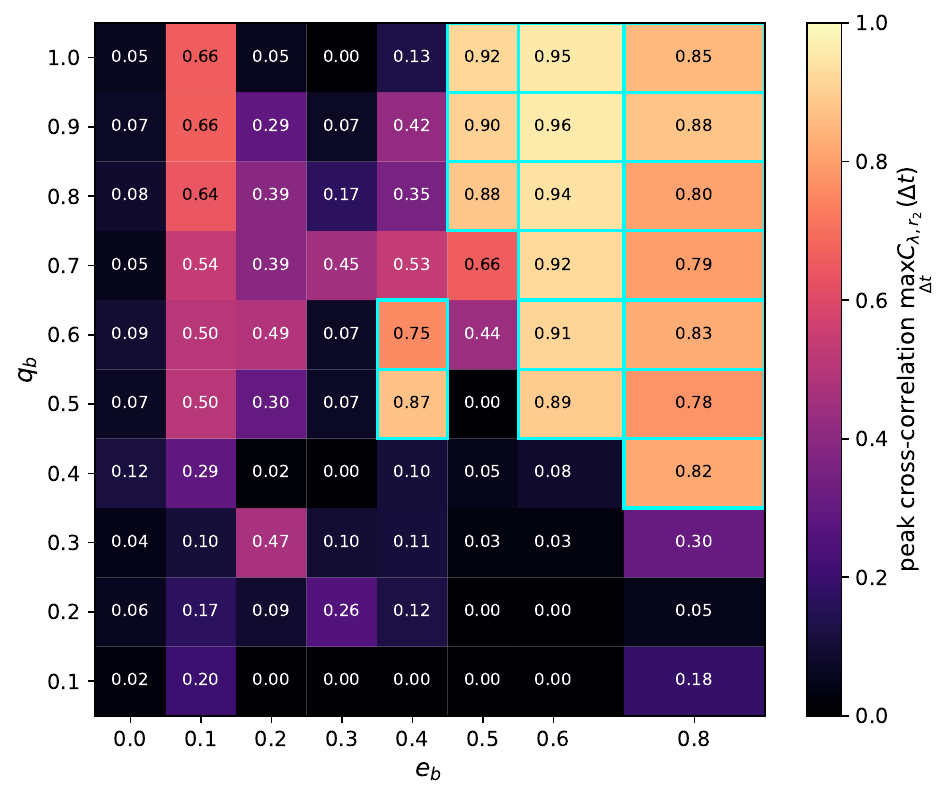}
    \caption{Peak cross-correlation $\max_{\Delta t} C_{\lambda, r_2}(\Delta t)$ between $\lambda(t)$ and the secondary--cavity-wall distance $r_2(t)$ across the $(e_b, q_b)$ suite (positive values only; see Appendix~\ref{sec:appendix_crosscorr}). Cyan outlines mark the strongly correlated cells ($\ge 0.7$). The strongly correlated cells coincide with the precessing, time-varying-$\lambda$ regime (\autoref{tab:stable_varying_grid}), supporting cavity-wall distance as the dominant pacing mechanism there.}
    \label{fig:cc_heatmap}
\end{figure}

\autoref{fig:cc_examples} shows $C_{\lambda, r_2}(\Delta t)$ for three representative simulations. In the two precessing cells, $(e_b, q_b) = (0.6, 1.0)$ and $(0.5, 0.9)$, the cross-correlation is a clean, large-amplitude curve that peaks at $C \approx 0.9$ for a small positive lag ($\Delta t \approx 30$--$50\,\tau_b$); in the messy cell $(0.4, 0.7)$ the curve is broad and weak. \autoref{fig:cc_heatmap} maps the peak correlation across the suite. The cells with strong correlation ($\max_{\Delta t} C_{\lambda, r_2} \ge 0.7$; cyan outlines) coincide with the precessing, time-varying-$\lambda$ cells of \autoref{tab:stable_varying_grid}, clustered at higher $e_b$. We read this as quantitative support for the cavity-wall distance pacing most of the preferential-accretion variability in the precessing regime, with the small but non-zero lag reflecting the finite response time of the accretion flow rather than the instantaneous proximity assumed by the naive picture of \S\ref{sec:preliminary_analysis}.

\bsp	
\label{lastpage}

\end{document}